\documentclass[
    aps,
    pra,
    twocolumn,
    longbibliography,
    floatfix,
    superscriptaddress
    ]{revtex4-1}
\usepackage{amsfonts}
\usepackage{amssymb}
\usepackage{amsmath}
\usepackage{amsthm}
\usepackage{bm}
\usepackage{braket}
\usepackage[export]{adjustbox}

\usepackage{graphicx}
\usepackage{dcolumn}

\usepackage[utf8]{inputenc} 

\usepackage[normalem]{ulem}
\usepackage{xcolor}
\usepackage[breaklinks=true]{hyperref}
\hypersetup{colorlinks=true,linkcolor=blue,citecolor=blue,filecolor=blue,urlcolor=blue,pdfstartview=FitH}

\definecolor{OliveGreen}{RGB}{0,100,0}

\def\ex{\mathbf{e}_x}  
\def\ey{\mathbf{e}_y}  
\def\ez{\mathbf{e}_z}  

\def\kr{k_{\mathrm R}}  
\def\ma{m_{\mathrm a}}

\begin{document}
\title{Stationary solitary waves in \texorpdfstring{$F=1$}{F=1} spin-orbit coupled Bose-Einstein condensates}

\author{T.~Mithun}
\affiliation{Department  of  Atomic and Molecular Physics,  Manipal Academy of Higher education, Manipal 576 104, India}
\affiliation{Department of Mathematics and Statistics, University of Massachusetts, Amherst MA 01003-4515, USA}

\author{A.~R.~Fritsch}
\affiliation{Joint Quantum Institute, National Institute of Standards and Technology,
and University of Maryland, Gaithersburg, Maryland, 20899, USA} 
\affiliation{Instituto de Física de São Carlos, Universidade de São Paulo, Caixa Postal 369, 13560-970, São Carlos, SP, Brazil}

\author{G.~N.~Koutsokostas}
\author{D.~J.~Frantzeskakis}
\affiliation{Department of Physics, University of Athens, Panepistimiopolis - Zografos, Athens 157 84, Greece, EU}

\author{I.~B.~Spielman}
\email{ian.spielman@nist.gov}
\homepage{http://ultracold.jqi.umd.edu}
\affiliation{Joint Quantum Institute, National Institute of Standards and Technology,
and University of Maryland, Gaithersburg, Maryland, 20899, USA} 

\author{P.~G.~Kevrekidis}
\email{kevrekid@umass.edu}
\affiliation{Department of Mathematics and Statistics, University of Massachusetts, Amherst MA 01003-4515, USA}

\begin{abstract}
We consider solitary wave excitations above the ground state of $F=1$ spin-orbit coupled Bose-Einstein condensates (SOBECs).
The low energy properties of SOBECs in any of the three branches of the single particle dispersion relation can be described by suitable scalar nonlinear Schrödinger (NLS) equations which we obtain using multiple-scale expansions.
This enables us to examine a variety of different configurations, such as dark solitary waves associated with higher energy branches, as well as dark and bright structures in the lowest branch. 
{The lowest branch can also exhibit a ``superstripe'' phase 
that supports solitary waves.}
In all cases, we provide explicit expressions for the NLS coefficients, and confirm their validity with full numerical simulations of the SOBEC system including a harmonic confining potential.
\end{abstract}
 
\maketitle


\section{Introduction}
Understanding the effect of synthetic spin-orbit coupling (SOC) in Bose-Einstein condensates (BECs) is an active topic in cold atom physics~\cite{galitski2013spin,lin2011spin,lin2009synthetic}. 
Starting from its first experimental realizations this topic has gained considerable traction~\cite{Sandro:15}, with the experimentally accessible case with equal contributions of Rashba~\cite{Bychkov1984} and Dresselhaus~\cite{Dresselhaus1955} SOC being,
arguably, the most studied.
The properties of spin-orbit coupled BECs (SOBECs) have been recently reviewed in Ref.~\cite{zhang_16} (with an emphasis on the so-called Dicke model and associated  phase-transitions).
While most of the relevant works have focused on two-component systems, prototypical higher spin cases have been proposed~\cite{wang2010spin} and realized~\cite{Campbell2016,Lan2014Raman}.

Most research on SOBECs has focused on systems in or near equilibrium, however, a number of studies have considered localized nonlinear excitations, i.e., solitary waves.
Early studies considered the dynamics of bright and dark solitary waves in 1D~\cite{PhysRevLett.110.264101,achilleos2013matter}; later work considered vortices and their ordering properties in 2D~\cite{galitski_14,fetter_14}; by now a progressively increasing body of work addresses such excitations~\cite{adhikari2020vortex,gautam2017vortex,guo2021dynamics,gautam2018three,gautam2015mobile,gautam2015vector,wang2010spin,zhu2020spin,mithun2019modulation,song2014fragmentation}.
More broadly, in 1D bright and dark solitons play a central role in the dynamics of atomic BECs~\cite{abdullaev,djf,Szankowski2010oscillating,Ieda2004exact}; as do vortices in two spatial dimensions~\cite{fetter}; and vortex lines and rings in three dimensions~\cite{komineas}.

Motivated by the growth of these areas and 
spinor condensates more generally~\cite{KAWAGUCHI2012253,stamper2013spinor}, the
present work considers solitary waves in higher spin SOBECs. 
Experimentally these could be realized using techniques developed earlier  by one of the present authors that 
were used in order to explore the ground states of such systems, giving ferromagnetic, polar and superstripe phases~\cite{Campbell2016}.
The collective excitation spectrum takes on the standard Bogoliubov form similar to the two-spin case~\cite{Ji2015}.
The present effort extends this analysis to the case of  solitary states that emerge in the
presence of mean-field nonlinearity in the vicinity of extrema in the single particle spectrum.
Our study uses multiscale expansions~\cite{jeffrey1982asymptotic,ablowitz2} to obtain closed-form (albeit approximate) descriptions of solitary wave 
excitations. We cross-check these against numerical solutions of the 1D Gross-Pitaevskii equation (GPE), a nonlinear Schrödinger equation describing isolated coherently evolving BECs.

Our analysis begins in Sect.~II where we establish the microscopic model and introduce the multiscale perturbation method.
We first validate the multiscale perturbation method by initially selecting parameters for which $F=1$ SOC is reminiscent of the well studied $F=1/2$ case.
In Sect.~III we obtain the linearized excitation spectrum (i.e., phonons), and controllably introduce nonlinearity, by expanding the solution in a power series of a parameter characterizing the departure from the linear limit.
The equations satisfied by the two leading order corrections identify an effective scalar nonlinear Schr{\"o}dinger (NLS) equation. 
The coefficients of the corresponding NLS model and their dependence on the linear and nonlinear system properties are explicitly computed.
Subsequently, in Sect.~IV, we evaluate these coefficients in each case of interest (near the extrema of the respective bands) giving both dark and bright solitary waves~\cite{achilleos2015positive,PhysRevLett.118.155301}.
More elaborate structures, including stripe-phase waves are also  considered.  
In Sect.~V, the results of all the cases are compared with direct numerical computations, both with and without a  realistic parabolic trap. 
Lastly in Sect.~VI, we present results for parameters where $F=1$ and $F=1/2$ SOC differ qualitatively. 
In Sect.~VII we conclude and consider possible directions of future study.

\section{Model} \label{sec:model}

\begin{figure}[tb!]  
  \includegraphics{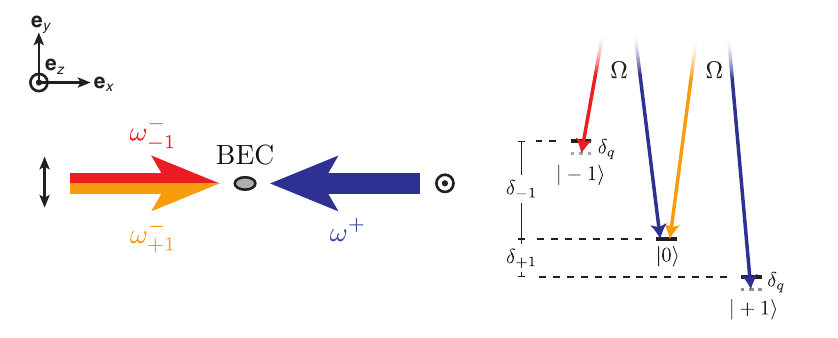}
\caption{
High magnetic field level diagram for the $^{87}$Rb $F=1$ hyperfine ground state when the atomic quadratic Zeeman shift is large, giving energy differences $\delta_{\pm1}$ between $\ket{F=1, m_F = \pm 1}$ and $ \ket{F=1, m_F = 0}$.
A pair of counter propagating laser beams, with polarization noted by black arrows, independently Raman couple the $\ket{F=1, m_F = -1}\leftrightarrow \ket{F=1, m_F = 0}$ and $\ket{F=1, m_F = 0}\leftrightarrow \ket{F=1, m_F = +1}$ transitions as was done experimentally in Ref.~\cite{Campbell2016}.
}
\label{fig:Setup}
\end{figure}

We consider quasi 1D spinor BECs with total angular momentum $F=1$ with SOC induced by Raman-coupling~\cite{lin2011spin} the three spin components $|m_F=0, \pm 1\rangle$ of the $F=1$ hyperfine ground state. 
The BEC, with typical per-particle interaction energy $\varepsilon$, is confined in a highly anisotropic trap with longitudinal and transverse frequencies, $\omega_x$ and $\omega_{\perp}$, respectively, such that $\hbar\omega_x,\varepsilon \ll \hbar\omega_{\perp}$.
The system can therefore be described by the 1D many-body Hamiltonian
\begin{align}
\hat {\mathcal H} =& \hat {\mathcal H}_{\rm sp} + \hat {\mathcal H}_{\rm int} \nonumber \\
=& \int d x \sum_{l, m} \hat \psi^\dagger_{l}(x) H_{l,m}(x)  \hat \psi_{m}(x) \label{eq:second_quantized_H}\\
&+ \frac{1}{2} \int d x:\left[ g_0 \hat n^2(x)+ \frac{g_2}{ \hbar^2}\left| \hat{\boldsymbol{\mathcal F}}(x) \right|^2 \nonumber
\right]:
\end{align}
where $:\!\cdots\!:$ denotes the normal ordering operation.
The Hamiltonian can be divided into a two-field single-particle term quantified by the single particle Hamiltonian operator $\hat H(x)$  with matrix elements $H_{l,m}(x)$ and a pair of four-field interaction terms with  ``symmetric'' (spin-independent) and ``antisymmetric'' (spin-dependent) interaction coefficients $g_0$ and $g_2$ respectively~\cite{nistazakis2008bright}.
These interaction constants are related to $a_0$ and $a_2$, the $s$-wave scattering lengths of two atoms with total spin $F=0$ and $F=2$, via $g_0=2\hbar \omega_{\perp}(a_0+2a_2)/3$ and 
$g_2=2\hbar \omega_{\perp}(a_2-a_0)/3$.

Here $\hat \psi^\dagger_{m}(x)$ describes the creation of a boson at position $x$ in magnetic sub-level $m_F = m$; 
\begin{align}
\hat n(x) &\equiv \sum_m \hat n(x) = \sum_m  \hat \psi^\dagger_m(x)  \hat \psi_m(x) \label{eq:density}
\end{align}
is the local density operator; and
\begin{align}
\hat{\boldsymbol{\mathcal F}}(x) &= \sum_\nu \left[\sum_{l, m} F^{(\nu)}_{l, m} (x)  \hat \psi^\dagger_{l}(x)  \hat \psi_{m}(x)\right] {\bf e}_\nu
\end{align}
with $\nu \in \left\{x,y,z\right\}$, is the angular momentum density vector operator in terms of the three $F=1$ angular momentum matrices $\hat F^{(\nu)}$.

We focus on a specific experimentally realized case shown in Fig.~\ref{fig:Setup} in which the $\ket{m_F = -1}\leftrightarrow \ket{m_F = 0}$ and $\ket{m_F = 0}\leftrightarrow \ket{m_F = +1}$ transitions are independently Raman coupled.
As derived in Appendix~\ref{app:Hamiltonian} we make a pair of rotating-wave approximations (RWAs) that lead to the single particle SOC Hamiltonian \cite{Zhai_2015,Goldman_2014,Campbell2016}
\begin{align}
\frac{\hat{H}_{0}}{\hbar} &\!=\!
  \begin{bmatrix}
    \frac{\hbar( -i \partial_x + k_R)^2}{2 \ma}\!+\!\delta_{\rm q} &  \Omega/2 & 0 \\
    \Omega/2 & -\frac{\hbar\partial_x^2}{2 \ma} & \Omega/2 \\
       0  & \Omega/2  & \frac{ \hbar(-i \partial_x- k_R)^2}{2 \ma}\!+\!\delta_{\rm q} \\
          \end{bmatrix},
\label{matrixH0}  
\end{align}
with the momentum operator $p = -i\hbar\partial_x$ and the atomic mass $\ma$.
The SOC Hamiltonian is additionally characterized by the wavenumber of the Raman coupling laser $k_R$, the Raman coupling strength $\Omega$, and a experimentally tunable parameter $\delta_{\rm q}$ analogous to the quadratic Zeeman shift.
In terms of the angular momentum operators 
$\hat F_{x,y,z}$, the total single particle Hamiltonian becomes
\begin{equation}
\hat{H}=\frac{(-i\hbar\partial_x \hat I + k_R \hat F_z)^2}{2\ma} + \frac{\delta_{\rm q}}{\hbar} \hat F_z^2 + \frac{1}{\sqrt{2}}\Omega \hat F_x + V(x) \hat I,
\label{eq:H0}
\end{equation}
where we included the spin-independent confining potential $V(x)=\ma \omega_{x}^2 x^2/2$ with the identity operator $\hat I$.
We note that this Hamiltonian can be represented in other forms as well by an appropriate pseudo-spin rotation. 
For example, one finds cross-terms that correspond to an equal weight to Rashba ($p_x \hat F_x + p_y \hat F_y$) and Dresselhaus ($p_x \hat F_x - p_y \hat F_y$) coupling \cite{Zhai_2015,Goldman_2014}.
At the same time, recent studies have considered pure Rashba coupling \cite{Goldman_2014,gautam2017vortex,Meng2016,Valdes-Curiel2021}.

The transformations leading to the single particle SOC Hamiltonian also modify the spin-dependent interaction energy that results from the angular momentum density (see App.~\ref{app:Hamiltonian} for details).
In terms field operators this takes the explicit form
\begin{align*}
:\left| \hat{\boldsymbol{\mathcal F}}(x) \right|^2:  & = \Big[ \left(
\hat \psi_{+1}^\dagger \hat \psi_{+1}^\dagger \hat \psi_{+1} \hat \psi_{+1} 
+ \hat \psi_{-1}^\dagger \hat \psi_{-1}^\dagger \hat \psi_{-1} \hat \psi_{-1} \right) \\ 
&\ + 2 \Big(\hat \psi_{+1}^\dagger \hat \psi_{0}^\dagger \hat \psi_{+1} \hat \psi_{0} 
+ \hat \psi_{-1}^\dagger \hat \psi_{0}^\dagger \hat \psi_{-1} \hat \psi_{0} \\
&\ - \hat \psi_{+1}^\dagger \hat \psi_{-1}^\dagger \hat \psi_{+1} \hat \psi_{-1} \Big) \\
&\ +2 \left(\hat \psi_{0}^\dagger \hat \psi_{0}^\dagger \hat \psi_{+1} \hat \psi_{-1}  
 + \hat \psi_{+1}^\dagger \hat \psi_{-1}^\dagger \hat \psi_{0} \hat \psi_{0} \right)\Big] 
\end{align*}
that includes contributions to the density-density interaction strength (first three lines) and spin-changing collisions (last line).  
The spin-changing collision terms are eliminated by the rotating wave approximation and we introduce $:|\hat{\boldsymbol{\mathcal F}}_{\rm RWA}(x)|^2:$ as the combined operator without these terms.

This final approximation is valid when the per-particle spin-dependent interaction energy scale $\epsilon_2 = g_2 \langle n(x) \rangle$ is much smaller than the quadratic Zeeman shift $|\delta_{-1} - \delta_{+1}|$, i.e., $\epsilon_2 \ll |\delta_{-1} - \delta_{+1}|$.
For the parameters in Ref.~\cite{Campbell2016} this is easily satisfied with $\epsilon_2 \approx h\times 5\ {\rm Hz}$ and $|\delta_{-1} - \delta_{+1}| \approx h\times 100\ {\rm kHz}$.

\subsection{Gross-Pitaevskii equation}

Here we turn to the mean-field description of this system suitable for weakly interacting atomic BECs described by  the 1D GPE with mean field energy density
\begin{equation}
\mathcal{E}=\sum_{l,m=-1}^1 
\psi_l^{\ast}\hat{H}_{lm}\psi_m+\frac{g_0}{2}n^2+\frac{g_2}{2 \hbar^2}|\boldsymbol{\mathcal F}_{\rm RWA}|^2,
\label{en}
\end{equation}
total energy $E=\int_{\mathbb{R}} \mathcal{E} dx$, and atom number $\int_{\mathbb{R}}n(x,t)dx=N$.
The density $n=\sum_{m=-1}^1|\psi_m|^2$ and RWA angular momentum density $|{\boldsymbol{\mathcal F}}_{\rm RWA}(x)|^2$ are the complex field analogues to the many-body quantities in Sec.~\ref{sec:model}.

We adopt dimensionless expressions with energy, length, time and density in units of $\hbar \omega_{\perp}$, $a_{\perp}$, $\omega_{\perp}^{-1}$ and $\sqrt{N/a_{\perp}}$ leading to dimensionless interaction coefficients $c_{0,2} \equiv g_{0,2} / (\hbar \omega_{\perp} a_{\perp})$ and the three-component GPE 
\begin{subequations}
\begin{eqnarray}
i\frac{\partial\psi_{+1}}{\partial t}&=&(\mathcal{L}+\Delta-i\gamma \partial_x)\psi_{+1}
+\Omega \psi_0 \label{eq: GPd1} \\
&+&c_2\left[
(\psi_{+1}^{\ast}\psi_{+1}+\psi_0^{\ast}\psi_0-\psi_{-1}^{\ast}\psi_{-1})\psi_{+1}\right], \nonumber \\
i \frac{\partial\psi_0}{\partial t}&=&\mathcal{L}\psi_0+ \Omega (\psi_{+1}+\psi_{-1})
\label{eq: GPd2} \\
&+& c_{2}(\psi_{+1}^{\ast}\psi_{+1}+\psi_{-1}^{\ast}\psi_{-1})\psi_0,
\nonumber \\
i\frac{\partial\psi_{-1}}{\partial t}&=&(\mathcal{L}+\Delta+i\gamma 
\partial_x)\psi_{-1}+\Omega \psi_{0}
\label{eq: GPd3} \\
&+&c_{2}\left[
(\psi_{-1}^{\ast}\psi_{-1} + \psi_0^{\ast}\psi_0-\psi_{1}^{\ast}\psi_{1})\psi_{-1}\right],
\nonumber
\end{eqnarray}
\end{subequations}
where
\begin{equation}
\mathcal{L}=\left[-\frac{1}{2}\partial_x^2+V(x)\right]
+c_0(\psi_{-1}^{\ast}\psi_{-1}+\psi_0^{\ast}\psi_0+\psi_{1}^{\ast}\psi_{1}),
\label{lop}
\end{equation}
and
\begin{equation}
\Delta =  \delta_{\rm q}+\frac{\gamma^2}{2}.
\label{Delta}
\end{equation}
In our units, the trapping potential becomes $V(x)=\lambda_t^2 x^2/2$, with $\lambda_t=\omega_x/\omega_{\perp} \ll 1$. 
Finally,  in the equations of motion we  introduced
$\gamma =a_\perp k_R$, and made the substitutions $\Omega \mapsto \Omega/(2\omega_{\perp}$), 
and $\delta_{\rm q} \mapsto \delta_{\rm q}/\omega_{\perp}$.  
In addition, we introduce the ratio $\beta=c_2/c_0$ which is $\beta=0.04$ for $^{23}$Na and $\beta=-0.0046$ for $^{87}$Rb~\cite{Klausen_2001,Kempen_2002,stamper2013spinor}.

In the following analysis, we consider the case of a symmetric linear energy spectrum with 
$\gamma=1$.
Finally, in our analysis and simulations, we restrict $\Omega$ to be in the interval $[0,6]$ and fix $\beta=-0.0046$.  
Having presented the lay of the land, we now turn to our analytical considerations for the associated model.

\subsection{Multiscale perturbation method}

We employ an analytical approach, similar to the one used in the case of binary SOBECs 
\cite{achilleos2013matter,achilleos2015positive,PhysRevA.89.033636}, to derive approximate solitary
solutions of the GPE Eqs.~(\ref{eq: GPd1})-(\ref{eq: GPd3}). In particular, we will use 
a multiscale perturbation method \cite{jeffrey1982asymptotic,ablowitz2} to derive an effective 
single-component GP equation; the latter supports exact dark and bright soliton solutions 
(in the absence of the trap), which are then used for the construction of approximate solitary wave solutions of the original model.  These will be tested against
direct numerical computations of stationary solutions of the full SOC equations.

\begin{figure*}[tbp]  
\includegraphics[left]{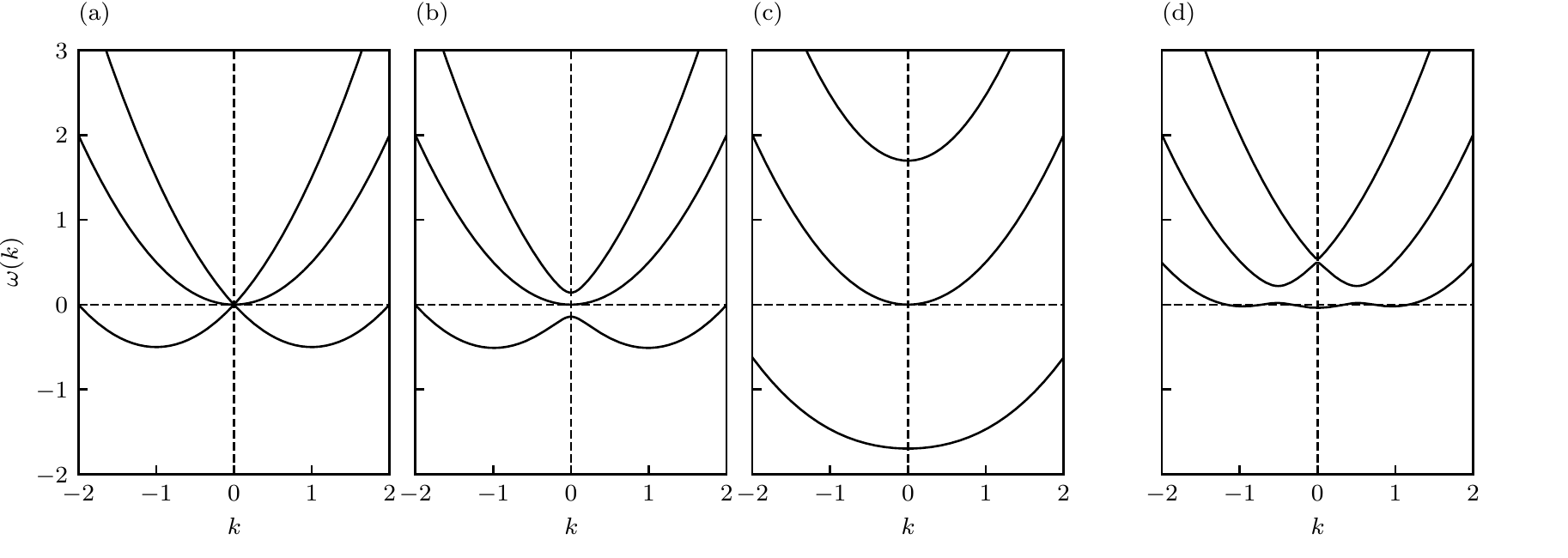}
\caption{  Single particle spectra for
$\Omega=0$ (a), $\Omega=0.1$ (b) and $\Omega=1.2$ (c) for the case of fixed $\Delta=0$ ($\delta_{\rm q} = -\gamma^2 / 2$) and  $\Omega=0.1$ (d) for  fixed $\Delta=\gamma^2 / 2$.
In the first three panels (a-c), 
lower, middle and upper lines respectively correspond to the
$\omega_1(k)$, $\omega_2(k)$ 
and $\omega_3(k)$,  of the dispersion relation of Eqs.~(\ref{b1})--(\ref{eq:band}); while, in the fourth panel the three lines represent three different bands of the single particle spectra computed by using Eq.~\eqref{matrixW0m}.
}
\label{fig:spectra}
\end{figure*}


First, we introduce the order parameter 
\begin{align}
\mathbf{\Psi} &= \mathbf{u} \exp[i (k x -\mu t)], & {\rm with} && \mathbf{u} &=(\phi_{1},\phi_{0},\phi_{-1})^{T}
\label{ban}
\end{align}
for an excitation with wavevector $k$, where the chemical potential $\mu=\omega+\epsilon^2 \omega_0$ governs the ground-state time-dependence.
$\omega$ represents the energy in the linear regime,  
while $\epsilon^2 \omega_0$ is a small deviation about this energy (with $0<\epsilon \ll1$ 
being a formal small parameter), and $\omega_0/\omega = O(1)$. Note that, as we will see below, 
$\omega_0$ will be a free parameter of the solutions. 
In the present context, we are seeking solutions that are bifurcating from the band edge
of the system's linear eigenstates.
Furthermore, we assume that the 
trapping potential is sufficiently weak, so that the normalized trap frequency is 
$\lambda_t = \epsilon^2 \tilde{\lambda}_t$.

Next, we introduce the following asymptotic expansions in $\epsilon$ for the fields $\phi_{m}$, 
with $m$ representing the magnetic quantum number, $m=(-1,0,1)$:
\begin{equation}
\phi_m=\sum_{i=1}^{\infty} \epsilon^i \phi_{mi},
\end{equation}
where the unknown fields $\phi_{mi}$ depend on the slow variables (since $\epsilon \ll 1)$
$$X=\epsilon x, \quad T=\epsilon^2 t.$$
Introducing the above ansatz into Eqs.~(\ref{eq: GPd1})-(\ref{eq: GPd3}), we arrive 
at the following equations at the orders $\mathcal{O}(\epsilon^1)$, $\mathcal{O}(\epsilon^2)$ 
and $\mathcal{O}(\epsilon^3)$, respectively
\begin{subequations}
\begin{align}
  \mathbf{W} \mathbf{u}_1 &= 0, \label{u1} \\
  \mathbf{W} \mathbf{u}_2 &= i \mathbf{W}_0 \partial_X\mathbf{u}_{1}, \label{u2} \\
  \mathbf{W} \mathbf{u}_3 &= i \mathbf{W}_0 \partial_X\mathbf{u}_{2}
\notag \\
  & \quad +\left(i\partial_T+\frac{1}{2}\partial_X^2
  -\mathbf{A}+\omega_0\right)\mathbf{u}_1,
\label{u3}
\end{align}
\label{eq:msp}
\end{subequations}
\hspace{-0.6em} 
where $\mathbf{u}_i=(\phi_{1i},\phi_{0i},\phi_{-1i})^{T}$ , while the matrices 
$\mathbf{W}$, $\mathbf{W}_0$ and $\mathbf{A}$ are given by:
%
%
\begin{subequations}
\begin{eqnarray}
\small \mathbf{W} & =&
\label{matrixW0}  
(\frac{k^2}{2}-\omega)\mathbf{I}+
  \begin{bmatrix}
    k\gamma+\Delta& \Omega& 0 \\
    \Omega & 0 &\Omega \\
       0  & \Omega  & -k\gamma+\Delta \\
  \end{bmatrix},
   \\
\label{matrixW}  
  \mathbf{W}_0 &=& \partial_k
  (\mathbf{W}+\omega \mathbf{I}),
  \\
  \mathbf{A} & =& \mathrm{diag}(a_1,a_2,a_3),
  \end{eqnarray}
\end{subequations}  
%
where $a_1$, $a_2$ and $a_3$ are given by:
\begin{subequations}
\begin{eqnarray}
a_1&=&c_0 n_t+c_2(-n_{-11}+n_{11}+n_{01})+\tilde{V}(X), 
\\
a_2&=&c_0 n_t+c_2(n_{11}+n_{-11})+\tilde{V}(X), 
\\
a_3&=&c_0 n_t+c_2(n_{-11}-n_{11}+n_{01})+\tilde{V}(X),
\end{eqnarray}
\end{subequations}
%
also $n_t=\sum_{m=-1}^1|\phi_{m1}|^2$, $n_{m1}=|\phi_{m1}|^2$, 
and the potential is given by  $\tilde{V}(X)=(1/2)\tilde{\lambda}_t^2 X^2$.

Equations~(\ref{eq:msp}) are a central finding of our  multiscale expansion method, and
can be used to obtain the results that follow.

\section{Analytical results for \texorpdfstring{$\Delta=0$}{Delta=0}}
In the subsequent analysis, we will set $\Delta=0$; 
this corresponds to a quadratic Zeeman shift $\delta_{\rm q}=-\gamma^2/2$.

\subsection{Linear regime}
First, at the leading order $O(\epsilon)$, which is relevant to the linear regime of the problem, 
we obtain the single particle energy spectrum $\omega(k)$. Indeed, the solvability condition 
$\text{det}\mathbf{W} = 0$ of Eq.~(\ref{u1}) yields three different branches, a lower, 
a middle and an upper one, namely $\omega(k)=\omega_j(k)$ ($j=1,2,3$), given by:  
\begin{subequations}
\begin{eqnarray}
\!\!\!\!\!\!
\omega_1&=&\frac{1}{2}(k^2-2\sqrt{k^2\gamma^2+2\Omega^2})~~(\text{lower branch}), 
\label{b1} \\
\!\!\!\!\!\!
\omega_2&=&\frac{k^2}{2}~~~~~~~~~~~~~~~~~~~~~~~~~~~~~(\text{middle branch}),
\label{b2} \\
\!\!\!\!\!\!
\omega_3&=&\frac{1}{2}(k^2+2\sqrt{k^2\gamma^2+2\Omega^2})~~(\text{upper branch}).
\label{b3}
\label{eq:band}
\end{eqnarray}
\end{subequations}
These branches of the energy spectrum are illustrated in Fig.~\ref{fig:spectra} for 
different values of the parameter $\Omega$. 
It is observed that (when the branches are separated), the upper and middle branches $\omega_2$ and $\omega_3$ 
feature a global minimum at $k=0$ 
for every value of $\Omega$. On the other hand, for $\Omega >\gamma^2/\sqrt{2}$, the lower branch 
$\omega_1(k)$ features a global minimum at $k=0$, while for 
$\Omega < \gamma^2/\sqrt{2}$ this branch acquires a double-well shape. 
In this case, 
$\omega_1(k)$ features a maximum at $k=0$ and two minima at 
$k=\pm \sqrt{\gamma^2-2(\Omega^2/\gamma^2)}$. The latter is, arguably, the
richest scenario in terms of relevant possibilities for solitary waveforms,
as we will illustrate below.

It is also straightforward to find that the solvability 
condition, $\text{det}(\mathbf{W})=0$, of Eq.~(\ref{u1}) leads to the solution 
\begin{equation}
\mathbf{u}_1 = \mathbf{R}\varphi(X,T),
\end{equation}
where $\varphi(X,T)$ is an unknown scalar field (to be determined below), while 
$\mathbf{R}=[Q_1,Q_2,Q_3]^T$ 
is the right eigenvector of the kernel of $\mathbf{W}$.
The components of $\mathbf{R}$ acquire different expressions for each branch of the energy spectrum. 

Next, we consider the equation at $O(\epsilon^2)$, namely Eq.~(\ref{u2}). 
Generally, the solvability condition of the inhomogeneous equations arising at $O(\epsilon^j)$ 
for $j\geq 2$ is $\mathbf{L}\mathcal{F}_j\mathbf{R}=0$, where $\mathcal{F}_j$ is the 
right-hand side term at $O(\epsilon^j)$. Hence, the solvability 
condition of Eq.~(\ref{u2}) is $\mathbf{L}\mathbf{W}_0\mathbf{R}=0$, where 
$\mathbf{L}=[Q_1,Q_2,Q_3]$ is the left eigenvector of the kernel of the matrix $\mathbf{W}$. 
The above solvability condition fixes the value of $k$, which is given by:
\begin{equation}
k=\gamma~\frac{Q_3^2-Q_1^2}{Q_1^2+Q_2^2+Q_2^2}.
\end{equation}
At this value of $k$, the group velocity becomes zero, i.e., 
\begin{equation}
v_g\equiv \omega'(k) =k-\gamma~\frac{Q_3^2-Q_1^2}{Q_1^2+Q_2^2+Q_2^2}=0.
\end{equation}
According to this result, perturbative solutions can only be sought for at the extrema 
(minima or maxima) of $\omega(k)$, which occur at the ``stationary points''  
$(\omega_m, k_m)$, where $\omega_m=\omega(k_m)$.
Based on this we expect the perturbative solutions to be approximately valid near these points.

Furthermore, at this order, a solution of Eq.~(\ref{u2}) reads:
\begin{equation}
u_2=-i (\partial_k\mathbf{R}) (\partial_X \varphi(X,T)).
\end{equation}
It is relevant to indicate at this point that this is a single {\it inhomogeneous
solution} of the Eq.~(\ref{u2}) and the most general associated solution can
be constructed by appending to it the solution of the homogeneous problem,
although we will not pursue this avenue herein.

\subsection{Nonlinear regime}

We now proceed with the equation at $O(\epsilon^3)$, namely Eq.~(\ref{u3}). 
The solvability condition of this equation is $\mathbf{L}\mathcal{F}_3\mathbf{R}=0$ 
(where $\mathcal{F}_3$ is the right-hand side of Eq.~(\ref{u3})). Then, employing 
the form of the solutions for $u_1$ and $u_2$, the solvability condition of Eq.~(\ref{u3}) 
yields the following effective GP equation,  
\begin{equation}
i\varphi_T=\left[-\frac{1}{2}\omega''(k_m) \partial_X^2
+\tilde{V}(X) + g(k_m)|\varphi|^2-\omega_0 \right]\varphi,
 \label{eq:eqscalar}
\end{equation}
where the dispersion and nonlinearity coefficients, $\omega''(k_m)$ and $g(k_m)$, are given by: 
\begin{widetext}
\begin{subequations}
\begin{eqnarray}
\omega''(k_m)&=&1+2\frac{k (Q_1 Q_1' + Q_2 Q_2'+ Q_3 Q_3')+\gamma (Q_1 Q_1'-Q_3 Q_3')}{Q_1^2+Q_2^2+Q_3^2},  
\label{omdis}
\\ 
g(k_m)&=&c_0 \frac{(Q_1^2+Q_2^2+Q_3^2)^2 + \big(2 Q_2^2 (Q_1^2+Q_3^2) + (Q_1^2 - Q_3^2)^2\big) \beta}{Q_1^2+Q_2^2+Q_3^2}.
\label{g}
\end{eqnarray}
\end{subequations}
\end{widetext}
Here, $\omega''(k_m)=\omega''(k)\Big|_{k=k_m}$, $g(k_m)=g(k)\Big|_{k=k_m}$ 
(i.e., the functions $\omega''(k)$ and 
$g=g(k)$ are evaluated at the stationary point $k_m$, as defined above) 
and $c_2=\beta c_0$. Notice that the 
coefficients $\omega''(k_m)$ and $g$ take different values for the three different branches 
of the dispersion relation. Furthermore, the relative sign of these coefficients controls the type 
of the soliton that is supported by the effective GPE [Eq.~(\ref{eq:eqscalar})]. 
In particular, considering time-independent solutions, in the absence of the potential ($\tilde{V}(X)=0$), and for 
$\omega''(k_m)g(k_m)>0$, the NLS Eq.~(\ref{eq:eqscalar}) possesses a stationary dark solitary (DS) 
solution of the form:  
\begin{equation}
\varphi_{DS}(X)=\sqrt{\frac{\omega_0}{|g(k_m)|}} 
\tanh\bigg(\sqrt{\frac{\omega_0}{|\omega''(k_m)|}}X\bigg),
 \label{eq:ds}
\end{equation}
while for $\omega''(k_m)g(k_m)<0$, it possesses a stationary bright soliton (BS) solution:
\begin{equation}
\varphi_{BS}(X)=\sqrt{\frac{2 \omega_0}{|g(k_m)|}} 
\text{sech}\bigg(\sqrt{\frac{2 \omega_0}{|\omega''(k_m)|}}X\bigg).
 \label{eq:bs}
\end{equation}
It is of course relevant to note that these stationary solutions can, in principle, be boosted using the Galilean transformation of the obtained
NLS equation~\cite{ablowitz2}.
Hence, in terms of the original variables, the system of Eqs. \eqref{eq: GPd3} yields a solitary wave solution of the form
\begin{eqnarray}
\mathbf{\Psi}(x,t; \mathbf{R}) &\approx& \bigg[\epsilon \varphi_S(\epsilon x) \mathbf{R}(k_m)-i \epsilon^2 \mathbf{R}'(k_m)\partial_X \varphi_S(\epsilon x) \bigg]
\nonumber \\ 
& \times& \exp[i (k_m x-\mu_m t)],
 \label{eq:dsappr}
\end{eqnarray}
valid to $O(\epsilon^3)$.
Here $\varphi_S$ is the (dark or bright) solution, and $\mu_m=\omega_m+\epsilon^2 \omega_0$. 
In light of the above
expression for Eq.~(\ref{eq:eqscalar}), the solitary wave mass will be inversely proportional to $\omega''(k_m)$, given the
nature of the contribution of the latter in the
equation's dispersive term; see also the details in Appendix A.

In general, there exist two different eigenfunction sets that we
consider herein (although different normalizations of the eigenvectors are
also possible; we comment on this a bit further below), labeled as $\mathbf{R_a}=[Q_{1a},Q_{2a},Q_{3a}]^T$ and $\mathbf{R_b}=[Q_{1b},Q_{2b},Q_{3b}]^T$, where 
\begin{equation}
\begin{split}
Q_{1a}(\omega,k)&=\frac{\bigg(\frac{k^2}{2}-k\gamma-\omega\bigg)}{\bigg(\frac{k^2}{2}+k\gamma-\omega\bigg)},\\
Q_{2a}(\omega,k)&=
-\frac{1}{\Omega}\bigg(\frac{k^2}{2}-k\gamma-\omega\bigg),~~~
Q_{3a}(\omega,k)=1
\label{eq:REF1}
\end{split}
\end{equation}
and
\begin{equation}
\begin{split}
 Q_{1b}(\omega,k)&=1,~~Q_{2b}(\omega,k)=
-\frac{1}{\Omega}\bigg(\frac{k^2}{2}+k\gamma-\omega\bigg),\\
Q_{3b}(\omega,k)&=1/Q_{1a}(\omega,k).
\label{eq:REF2}
\end{split}
\end{equation}
Since $\mathbf{R}_a$ and $\mathbf{R}_b$ are eigenfunctions, in line with earlier calculations in~\cite{achilleos2013matter,PhysRevLett.110.264101},
using a linear combination (which, by a continuation argument, may also exist in the nonlinear regime), we may also construct the solitary wave solution, 
which is of the form: 
\begin{eqnarray}
\mathbf{\Psi}(x,t) \approx \frac{C}{2} \Big[\Psi(\mathbf{R}_a) e^{ik_m x} +\Psi(\mathbf{R}_b) e^{-ik_m x}\Big],
\label{eq:dslc}
\end{eqnarray}
where $C$ is an arbitrary  constant.  For finite $k_m$, Eq.~(\ref{eq:dslc}) represents a stripe solitary wave solution.
This is in analogy with the stripe-phase ground state which contains density modulations resulting from interfering contributions to the mean-field wavefunction~\cite{wang2010spin,Ho2011,lin2011spin}.

Below we will present results for the type of solitary
wave that is supported at each branch of the 
energy spectrum, and corroborate our predictions with  results of direct numerical simulations.

\section{Solitary waves in a homogeneous BEC}

\subsection{Solitary waves at the lower branch}

First we consider the lower branch, $\omega_1(k)$, of the energy spectrum, which features either a single minimum at $k_m=0$ for $\Omega > \gamma^2/\sqrt{2}$, or 
a double-well shape with two minima  $k_m=\pm\sqrt{\gamma^4 - 2 \Omega^2}/\gamma$ for $\Omega < \gamma^2/\sqrt{2}$. In the following paragraphs, we discuss the solitary solutions for both the cases $\Omega > \gamma^2/\sqrt{2}$ and $\Omega < \gamma^2/\sqrt{2}$. 
\subsubsection*{
Case I: Dark solitary waves for \texorpdfstring{$k_m=0$, $\Omega > \gamma^2/\sqrt{2}$}{km=0, Omega > gamma\^2 / sqrt(2)}
}

In this case $\omega_m(k_m)= -\sqrt{2}\Omega$ (see Fig.~\ref{fig:spectra}(c)). Then, at $(\omega, k)=(\omega_m, k_m)$, we obtain:
 \begin{eqnarray}
 \mathbf{R}_a &=& [1,-\sqrt{2},1]^T, \quad 
 \mathbf{R}_a'=\left[-\frac{\sqrt{2}\gamma}{\Omega},
 \frac{\gamma}{\Omega},0\right]^T, 
\nonumber \\
\omega''(k_{m}) &=& 1-\frac{\gamma^2}{\sqrt{2}\Omega}, \quad g(k_{m})=2c_0(2+\beta),
 \nonumber
 \end{eqnarray}
 and, similarly, 
 \begin{eqnarray}
\mathbf{R}_b &=&[1,-\sqrt{2},1]^T, \quad 
\mathbf{R}_b'=\left[0,
 -\frac{\Omega}{\gamma},\frac{\sqrt{2}\Omega}{\gamma}\right]^T,
 \nonumber \\
 \omega''(k_{m}) &=& 1-\frac{\gamma^2}{\sqrt{2}\Omega}, \quad g(k_{m})=2c_0(2+\beta).
 \nonumber
\end{eqnarray} 
Observe that since $\omega''(k_{m}) >0$ and $g(k_{m})>0$
(for $c_0>0$), the stationary solution is a DS, as per Eq.\eqref{eq:ds}.
\begin{figure}[!htbp] 
\includegraphics{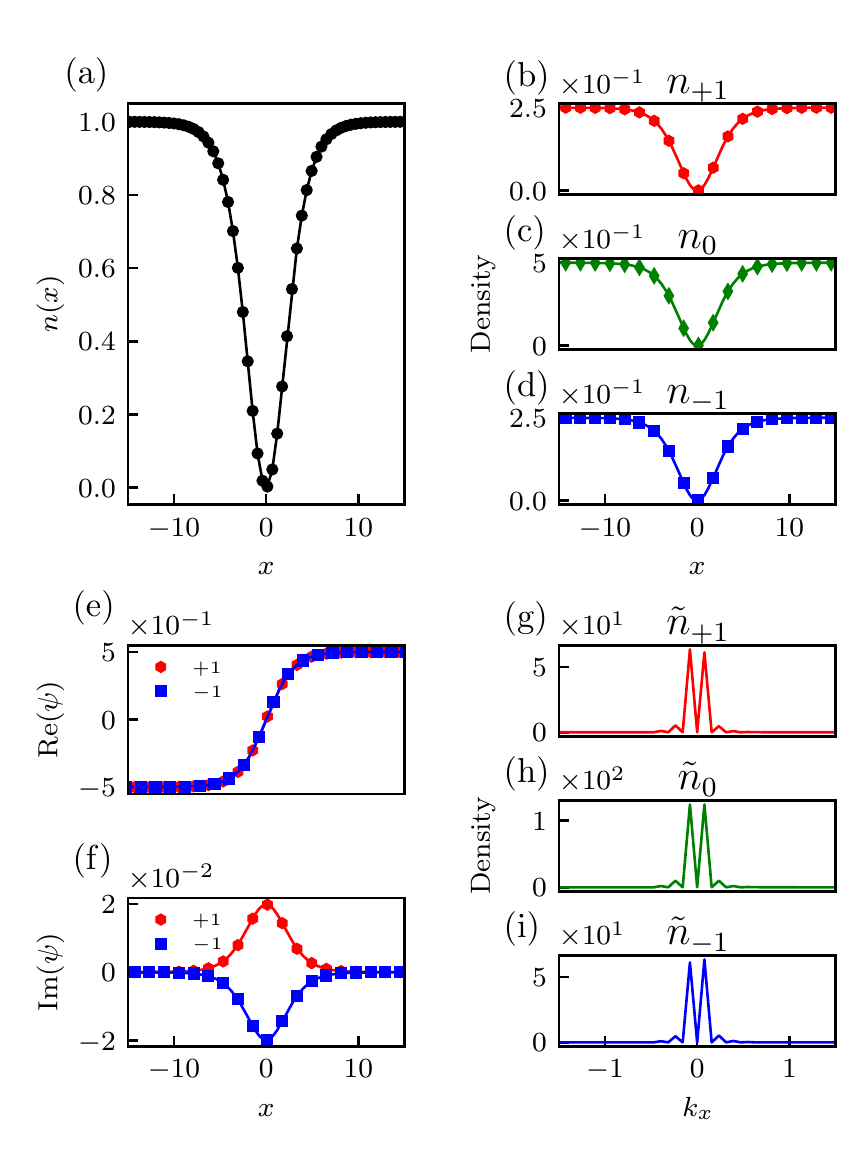}
\caption{The steady state DS solution for $k_m=0$ and $\Omega>\frac{\gamma^2}{\sqrt{2}}$. 
The panels depict: (a) The total density, (b)-(d) the density of individual components, (e) the real part of $(\psi_{+1},\psi_{-1})$, f) imaginary part of $(\psi_{+1},\psi_{-1})$, and (g)-(i) the density of individual components in Fourier space. The solid lines
represent numerical results, the symbols represent the theoretical prediction of (Eq.~\ref{eq:dslc}), with circles representing the total density $n$ and squares, diamonds and hexagrams denoting the spinor components with $m=-1$, $m=0$ and $m=+1$, respectively. The parameters are $c_0=1$, $\Omega=6$, $\gamma=1$, $\lambda_t=0$. $\mu=\omega_m+0.1$ and $C=1$.}
\label{fig:O_6_0}
\end{figure}

\begin{figure}[!htbp]  
\includegraphics[width=\columnwidth]{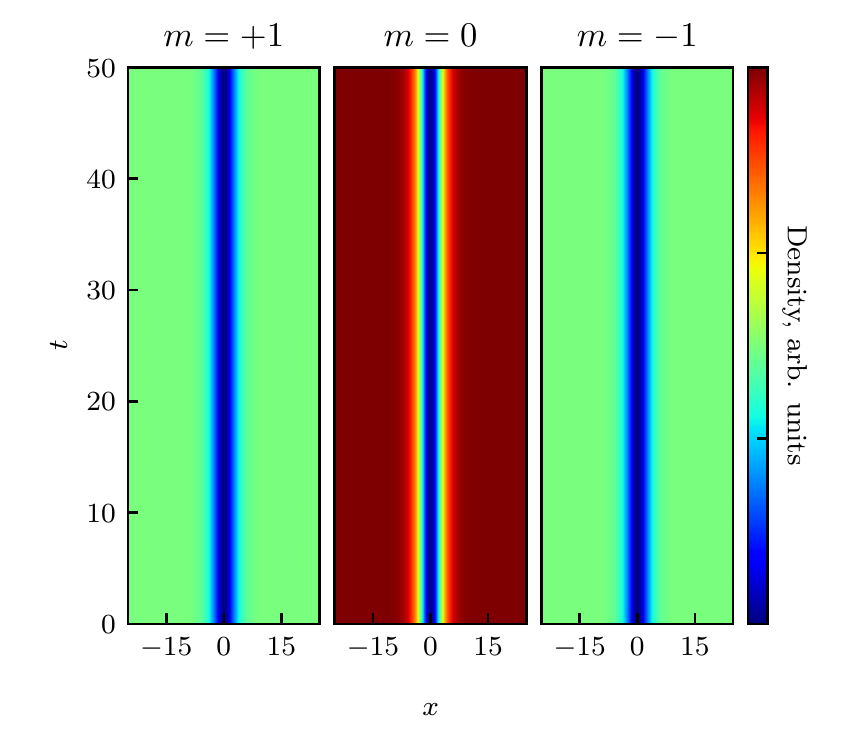}
  \caption{Density for a stationary DS as a function of time. The three spin components, $m=+1$ (left), $m=0$ (middle) and $m=-1$ (right) are shown. 
  The simulation parameters are $c_0=1$, $\Omega=6$, $\gamma=1$, $\beta=-0.0046$, $\lambda_t=0.0$, and $\mu=-8.38$.
}
\label{fig:DSevolvenotra}
\end{figure}

We now numerically solve the time-independent version of Eqs.~(\ref{eq: GPd1})-(\ref{eq: GPd3}) by considering $\mu_m=\omega_m+\epsilon^2 \omega_0$. The result of $\Omega=6$ at $\epsilon^2 \omega_0=0.1$ is shown in Fig.~\ref{fig:O_6_0}. We observe that the amplitudes of the dark solitary waves of the components $m=+1$ and $m=-1$  are equal, as obtained analytically in Eq.~\eqref{eq:dslc}. We further ensure that both the numerical and the analytical results {for the total density} 
are matching well by showing them on top left panel of Fig.~\ref{fig:O_6_0}. Further,
the figure shows that the real parts  of the wave functions $\psi_{+1}$  and $\psi_{-1}$ are equal; both the relevant
real and imaginary parts are shown by means of connected 
symbols in the left panels of the figure.
On the other hand, the change in sign of the profile of imaginary parts follows the analytical solution $u_2$. {In this manuscript, the density of the individual components in real space is normalized by the maximum of $n(x)$, and the real and imaginary parts of the wave functions normalized by the maximum of $\sqrt{n(x)}$.} The representation of the wave functions in the Fourier space shows the contributing momentum values.  Overall,
we confirm that the theoretical prediction adequately captures the numerically obtained solutions in all relevant components. Additionally, we confirmed the stability of this solution both by evolving it for a longer time as shown Fig.~\ref{fig:DSevolvenotra} and with a full stability analysis of the Bogolyubov-de Gennes (BdG) equations~\cite{frantzeskakis2015defocusing}.  For details of the corresponding stability calculation, see Appendix~\ref{sec:eig}.


 %
 \begin{figure}[tbp!]  
\includegraphics{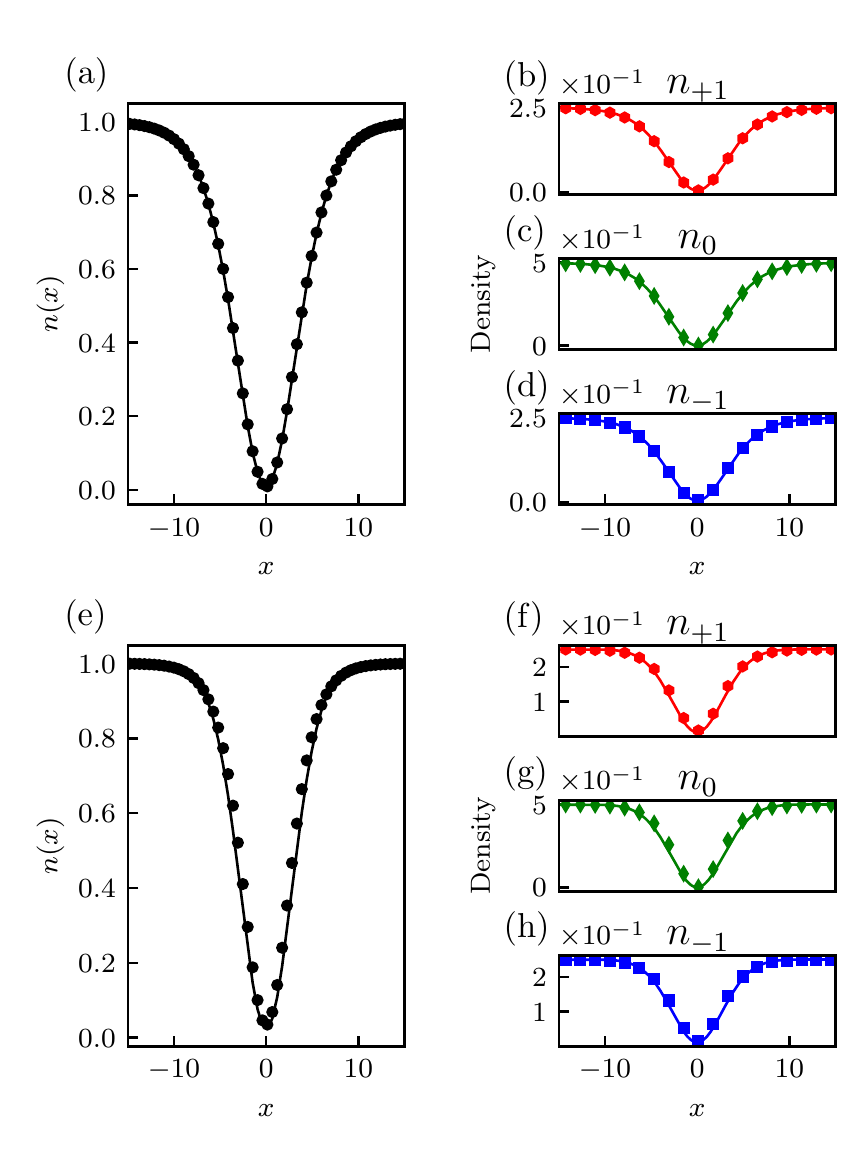}
\caption{The comparison between the theoretical and numerical results for ($\Omega=1.2$, $\epsilon^2 \omega_0=0.02$) (top panel) and ($\Omega=0.8$, $\epsilon^2 \omega_0=0.01$) (bottom panel).  
These parameters are still above the critical $\Omega$ below which a non-vanishing
$k_{min}$ exists. The panels depict: 
(a) and (e) The total density, and (b)-(d) and (f)-(h) the density of individual components. The solid line represents the numerical results and the symbols represent the analytical prediction (Eq.~(\ref{eq:dslc})). Here, circles represent total density $n$ and squares, diamonds and hexagrams denote spinor components with $m=-1$, $m=0$ and $m=+1$, respectively.}
\label{fig:O_1_2_0_8c}
\end{figure}

To corroborate that the DS solution exists and follows analytical predictions for all values of $\Omega > \frac{\gamma^2}{\sqrt{2}} $, we have considered two additional cases $\Omega=1.2$ and $\Omega=0.8$. The results are shown in Fig.~\ref{fig:O_1_2_0_8c}. As seen, the numerical results are in line with the analytical predictions, identifying a robust dark solitary wave where one such is expected to exist. 
\subsubsection*{Case II: Dark solitary waves for \texorpdfstring{$k_m=\pm\sqrt{\gamma^4 - 2 \Omega^2}/\gamma$, $\Omega < \gamma^2/\sqrt{2}$}{km=+- sqrt(gamma\^4 - 2 Omega\^2), Omega < gamma\^2 / sqrt(2)}}
\begin{figure}[tb!]  
\includegraphics{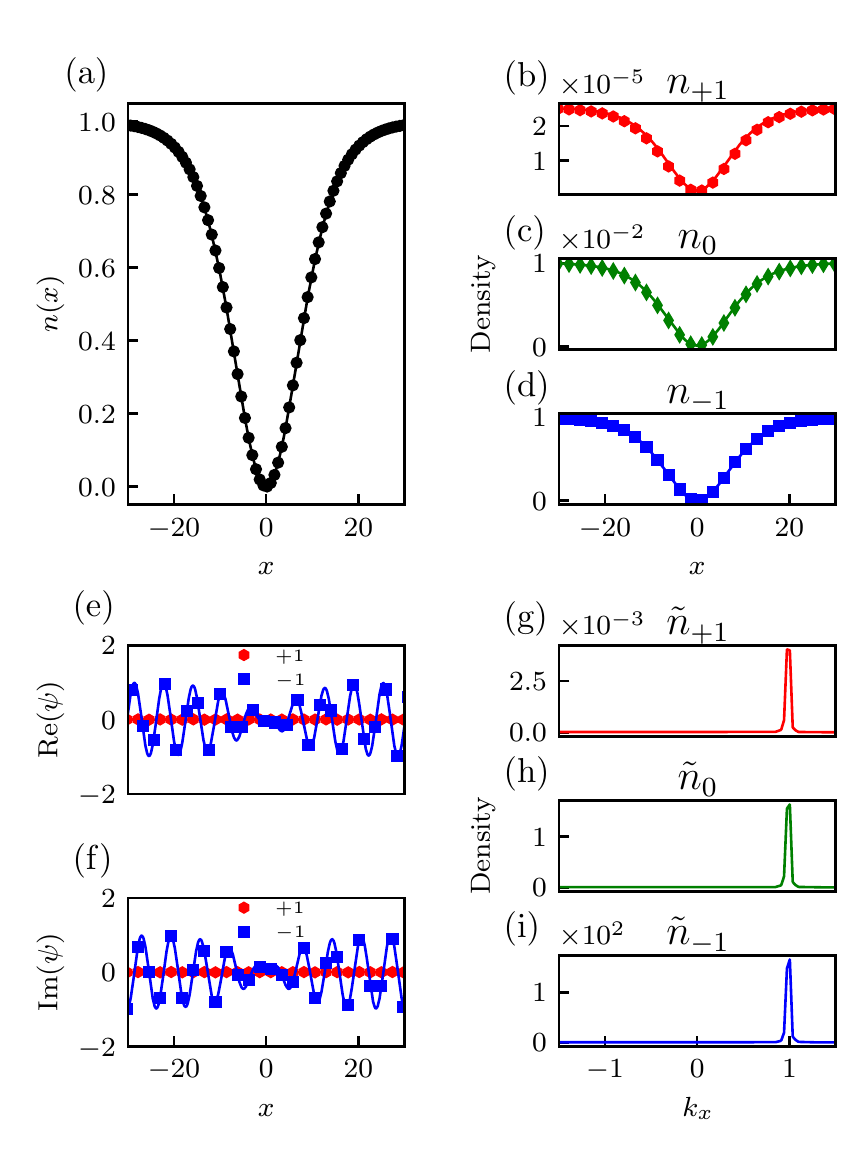}
\caption{
The steady state DS solution for the double-well case with $k_m=\sqrt{\gamma^4 - 2 \Omega^2}/\gamma$ and  $\Omega<\gamma^2/\sqrt{2}$. The panels depict: (a) The total density, (b)-(d) the density of individual components, (e) the real part of $(\psi_{+1},\psi_{-1})$, (f) the imaginary part of $(\psi_{+1},\psi_{-1})$, and (g)-(i) the density of individual components in Fourier space. The solid line represents numerical results and symbols represent the analytical prediction (Eq.~(\ref{eq:dsappr}) for $\mathbf{R}_a$).
Here, circles represent the total density $n$ and squares, diamonds and hexagrams denote spinor components with $m=-1$, $m=0$ and $m=+1$, respectively.  The parameters are $c_0=1$, $\Omega=0.1$, $\gamma=1$, $\lambda_t=0$ and $\mu=\omega_m+0.01$. }
\label{fig:km_0_1}
\end{figure}

\begin{figure}[!htbp]  
 \includegraphics[width=\columnwidth]{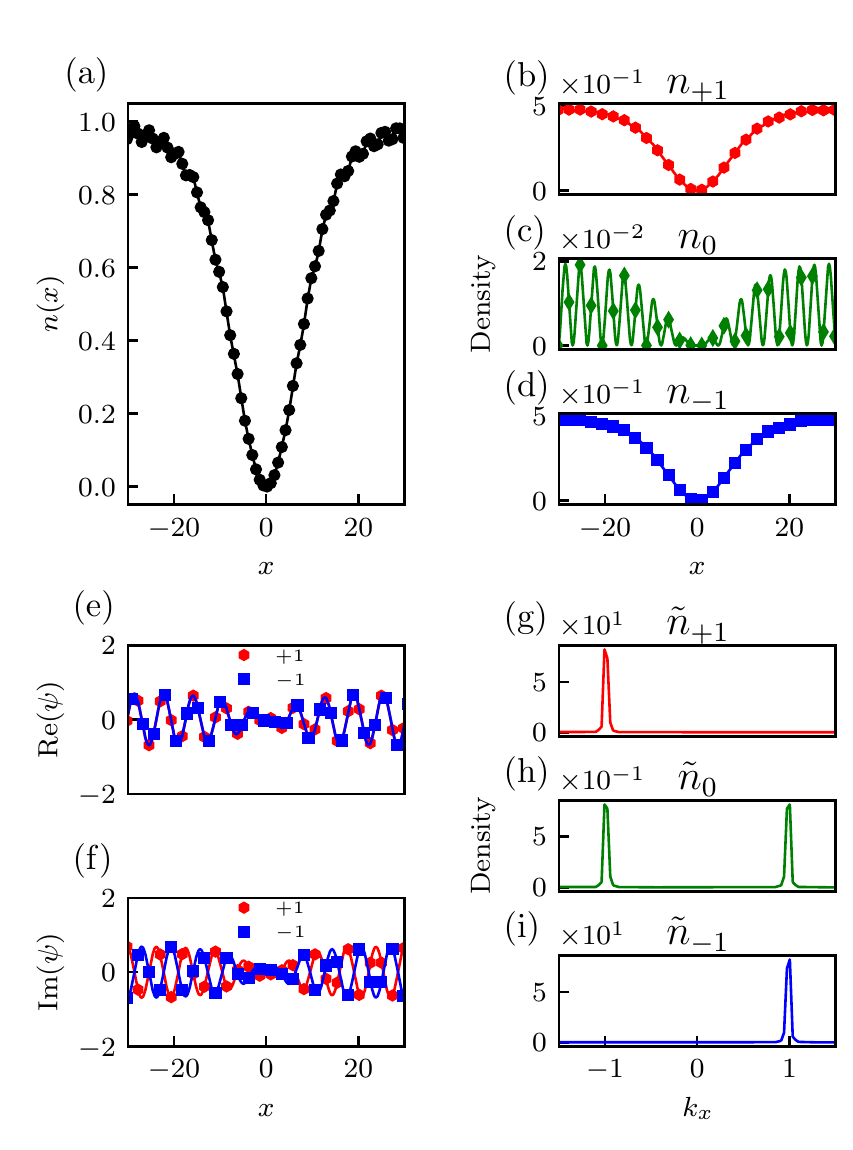}
\caption{
The stripe DS solution occurring at the double-well-shaped lower branch. 
The panels depict: (a) The total density, (b)-(d) the density of individual components, (e) the real part of $(\psi_{+1},\psi_{-1})$, (f) the imaginary part of $(\psi_{+1},\psi_{-1})$, and (g)-(i) the density of individual components in Fourier space. The solid line represents numerical results, and the symbols ---which have the same representation as before--- represent the analytical prediction (Eq.~(\ref{eq:dslc})). 
The parameters are $c_0=1$, $\Omega=0.1$, $\gamma=1$, $\lambda_t=0$. $\mu=\omega_m+0.01$ and $C=1.4$. }
\label{fig:km_0_1_ss}
\end{figure}

In this double-well case, $\omega_m(k_m)= -\frac{\gamma^4+2\Omega^2}{2\gamma^2}$ (see Fig.~\ref{fig:spectra}(b)). Then, at $(\omega, k)=(\omega_m, k_m)$, we obtain:
\begin{eqnarray}
\mathbf{R}_a &=&\left[\frac{- k_m +\gamma}{ k_m+\gamma},~\frac{ k_m \gamma-\gamma^2}{\Omega},~1\right]^T, 
\nonumber \\   
\mathbf{R}_a'&=&\left[\frac{2 (k_m -\gamma)}{\gamma(k_m + \gamma)},~\frac{- k_m + \gamma}{\Omega},~0\right]^T,
\nonumber \\
\omega''(k_{m})&=&1-\frac{2 \Omega^2}{\gamma^4},
\quad
 g(k_{m})=4 c_0\frac{(1+\beta)\gamma^8-2\beta \Omega^4}{\gamma^6 (k_m + \gamma)^2}.
\nonumber
 \end{eqnarray}
Similarly, in this case we construct:
\begin{eqnarray} 
\mathbf{R}_b &=& \left[1,~\frac{ - k_m \gamma-\gamma^2}{\Omega},~\frac{ k_m +\gamma}{- k_m+\gamma}\right]^T, 
\nonumber \\
\mathbf{R}_b'&=&\left[0,~\frac{ - k_m - \gamma}{\Omega},~\frac{2 (k_m + \gamma)}{\gamma(- k_m + \gamma)}\right]^T, 
\nonumber \\ 
\omega''(k_{m}) &=& 1-\frac{2 \Omega^2}{\gamma^4}, \quad g(k_{m})=4 c_0\frac{(1+\beta)\gamma^8-2\beta \Omega^4}{\gamma^6 (k_m -\gamma)^2},
\nonumber
\end{eqnarray}
Since, in this case too, $\omega''(k_{m}) >0$ and $g(k_{m})>0$, the stationary solution is again a dark solitary wave, per Eq.~\eqref{eq:ds}. Additionally, since $k_m=\pm\sqrt{\gamma^4 - 2 \Omega^2}/\gamma$, Eq.~\eqref{eq:dsappr} for $\mathbf{R}_a$ and $\mathbf{R}_b$ also provides solutions. The result corresponding to to the case $k_m=\sqrt{\gamma^4 - 2 \Omega^2}/\gamma$ is shown in Fig.~\ref{fig:km_0_1} where the dark solitary wave is depicted at the right momentum minimum. 
Naturally, there is a corresponding state around the left momentum minimum, with the relative populations of the $\psi_{+1}$ and $\psi_{-1}$ components reversed (not shown here
for brevity).
On the other hand, Fig.~\ref{fig:km_0_1_ss} shows the stripe solitary wave obtained from a linear combination of plane waves of momenta $\pm k_m$. It is interesting to note that 
despite the presence of a definitive finite wavenumber in the Fourier 
spectrum of the different components, the solution does not travel
due to its bifurcation from a point in $k$-space where the group velocity is vanishing.

Given that wavelength-scale spatial modulations visible in Fig.~\ref{fig:km_0_1} have a comparable length scale to the transverse confinement length $a_\perp$, it is relevant to briefly comment on the effective one-dimensionality of the system.
While length scales can be a useful heuristic, energies are the more suitable quantities to compare when identifying the validity of dimensional reduction.
Here, the interfering momentum components have energy of just $2 E_r \approx 7\ {\rm kHz}$
and the transverse confinement is generated by an optical lattice of depth $V = s E_r$~\footnote{To avoid introducing a lattice recoil energy $E_L$ we assume that the lattice is generated with lasers of wavelength close to that of the Raman lasers.
This is almost always accurate at the 30\ \% level since in alkali atoms these lasers all couple via nS to nP transitions.}.
In this case the vibrational spacing between the ground and first excited transverse states has energy $\approx 2 E_r \sqrt{s}$; for typical confining lattice depths of $s > 16 E_r$ this implies a spacing of $8 E_r \gtrsim 28\ {\rm kHz}$.
As such, coupling to these excited states is energetically blocked, confining transverse motion to the ground state.

\subsubsection*{Case III: Bright solitary wave for 
\texorpdfstring{$k_m=0$, $\Omega < \gamma^2/\sqrt{2}$}{km=0, Omega < gamma\^2 / sqrt(2)}
}

We additionally consider the case corresponding to the local maximum at $k_{m}=0$ of the double-well-shaped lower branch (occurring for $\Omega < \gamma^2 / \sqrt{2}$), for which  
$\omega_m(k_m)= -\sqrt{2}\Omega$ (see the left panel of Fig.~\ref{fig:spectra}). 
For this case, at $(\omega, k)=(\omega_m, k_m)$, we find:
\begin{eqnarray}
\mathbf{R}_a &=& [1,-\sqrt{2},~1]^T, \quad
\mathbf{R}_a'=\left[-\frac{\sqrt{2}\gamma}{\Omega},
 \frac{\gamma}{\Omega},~0\right]^T,
\nonumber \\ 
 \omega''(k_{m})&=&1-\frac{\gamma^2}{\sqrt{2}\Omega}, \quad
 g(k_{m})=2c_0(2+\beta), 
\nonumber 
\end{eqnarray}
and similarly,
\begin{eqnarray}
\mathbf{R}_b&=&[1,~-\sqrt{2},~1]^T \quad
\mathbf{R}_b'=\left[0,
~-\frac{\Omega}{\gamma},~\frac{\sqrt{2}\Omega}{\gamma}
\right]^T,
\nonumber \\
\omega''(k_{m})&=&1-\frac{\gamma^2}{\sqrt{2}\Omega}, \quad g(k_{m})=2c_0(2+\beta). 
\nonumber  
\end{eqnarray}
Here, an important observation is that, while the nonlinearity coefficient is positive, $g(k_m)>0$, the dispersion coefficient changes sign, since
$\omega''(k_{m}) <0$. The latter coefficient is connected with the inverse of the effective mass, i.e., 
$m_{eff} \propto 1/\omega''(k_m)$ (see, e.g., Refs.~\cite{obert1,obert2}), which suggests that the solitary waves in this case feature a negative effective mass; this result can also be obtained by employing symmetry considerations (see details in Appendix~\ref{meff}). In the case of all the
dark solitary waves that are presented in this work, the 
structures are characterized by a positive effective mass~\footnote{Here we select a sign convention yielding a positive mass.  Typically the energy of a dark soliton is a decreasing function of its velocity, which is often described as resulting from a negative inertial mass.}. 

\begin{figure}[tb!]  
  \includegraphics[width=\columnwidth]{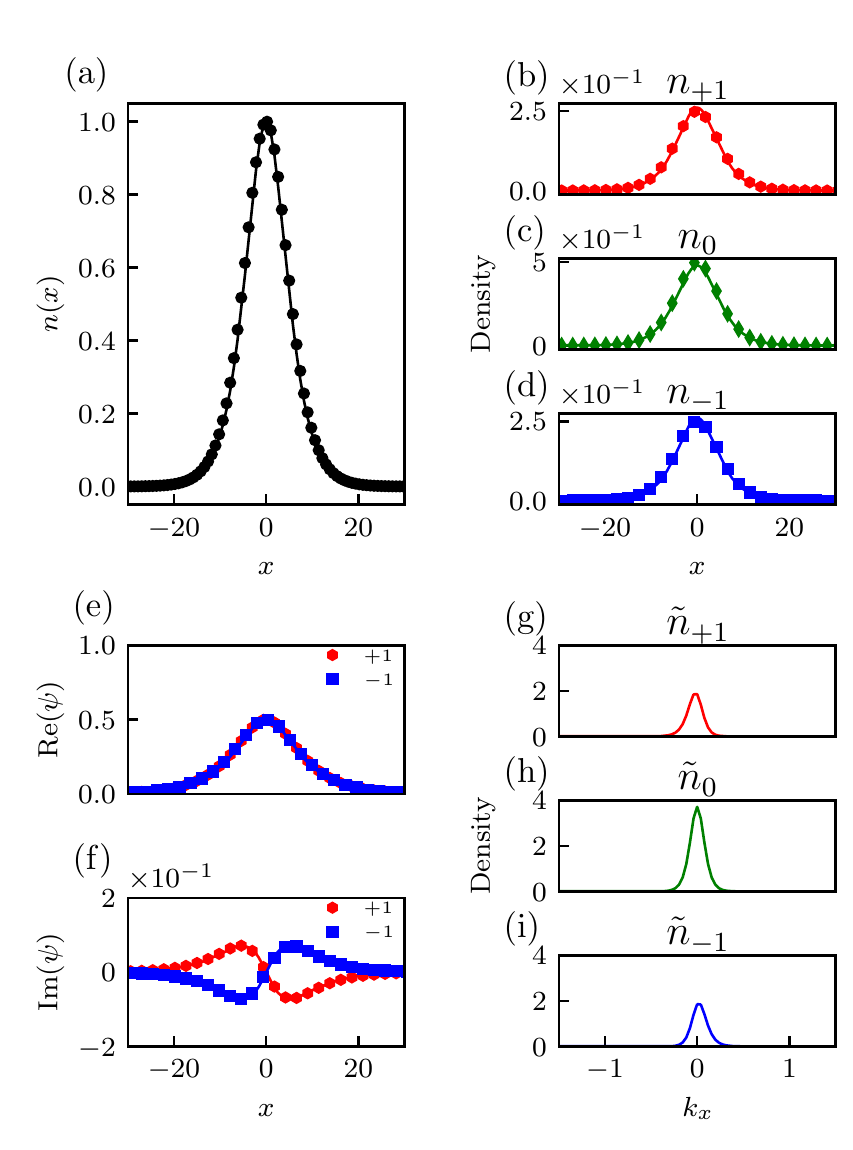}
\caption{The steady state BS solution for $\Omega<\gamma^2/\sqrt{2}$. a) The total density, b-d) the density of individual components, e) real part of $(\psi_{+1},\psi_{-1})$, f) imaginary part of $(\psi_{+1},\psi_{-1})$ and g-i) the density of individual components in Fourier space. The solid line represents numerical results and symbols represent theory (Eq.~\eqref{eq:dslc}), where circles represent total density $n$ and squares, diamonds and hexagram denote spinor components $m=-1$, $m=0$ and $m=+1$, respectively. The parameters are $c_0=1$, $\Omega=0.4$, $\gamma=1$, $\lambda_t=0$. $\mu=\omega_m+0.01$ and $C=1$. }
\label{fig:bs04}
\end{figure}

Importantly, since $g(k_{m})>0$ and $\omega''(k_{m}) <0$, the negative mass solitary wave is a {\it bright} one. The  functional form of this solitary wave is given by Eq.~(\ref{eq:bs}), and is illustrated in  
Fig.~\ref{fig:bs04}.  {We further confirmed that this solution is a spectrally stable coherent structure from the full stability analysis of the BdG equations.}
  
\subsection{Solitary waves at the middle branch}

\begin{figure}[!htbp]  
\includegraphics[width=\columnwidth]{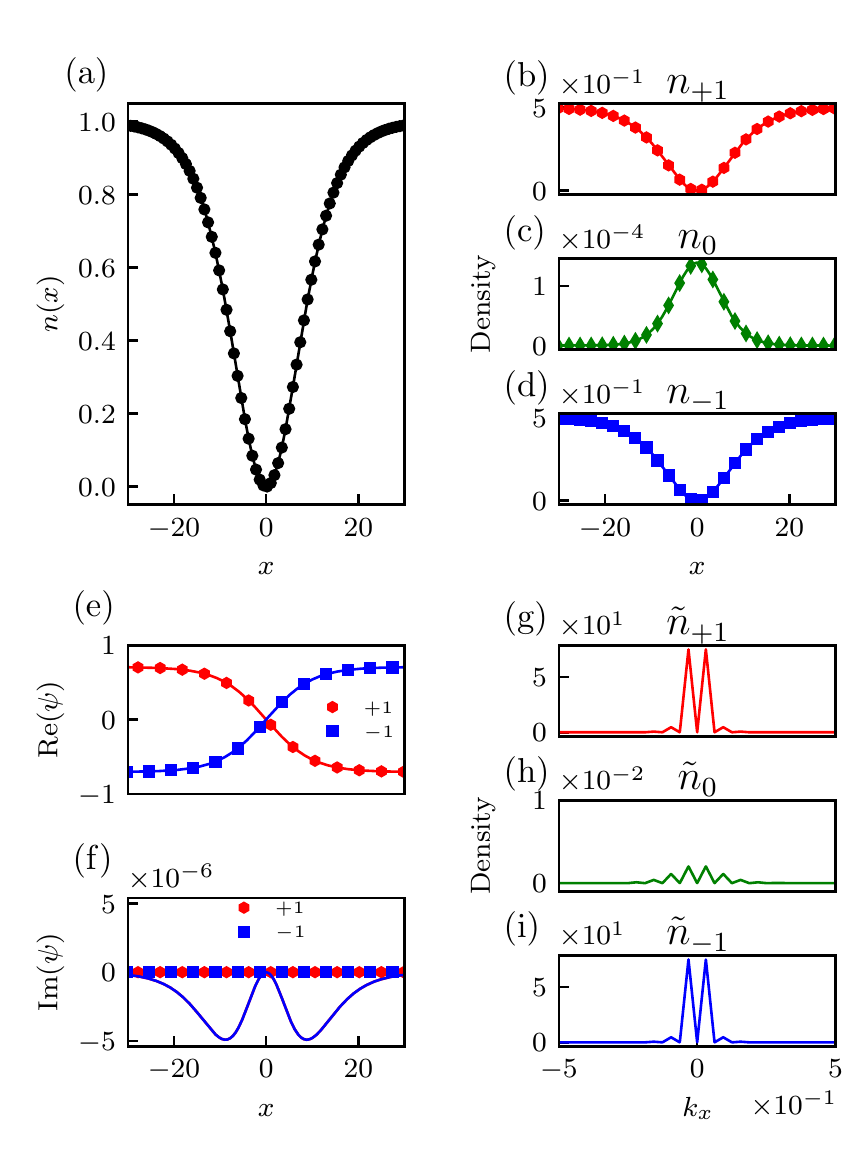}
\caption{The steady state DS solution for the middle branch. The panels depict: (a) The total density, (b)-(d) the density of individual components, (e) the real part of $(\psi_{+1},\psi_{-1})$, (f) the imaginary part of $(\psi_{+1},\psi_{-1})$, and (g)-(i) the density of individual components in Fourier space. The solid line represents numerical results and the symbols correspond
to the analytical approximation of Eq.~(\ref{eq:dsappr})  for $\mathbf{R}_a$.
The parameters are $c_0=1$, $\Omega=6$, $\gamma=1$, $\lambda_t=0$. $\mu=\omega_m+0.01$.
}
\label{fig:dsmb}
\end{figure}

We now consider structures that can be formed within 
the second branch of the dispersion relation $\omega_2(k)$. This branch has a minimum at $k_m=0$, while it is straightforward to  find that 
$\omega_m(k_m)= 0$; see Fig.~\ref{fig:spectra}. Then, we obtain the corresponding eigenvectors associated with the first and second order
 solutions:
\begin{eqnarray}
\mathbf{R}_a&=&[-1,~0,~1]^T, \quad
 \mathbf{R}_a'=\left[0,
 ~\frac{\gamma}{\Omega},~0\right]^T,
 \nonumber \\
\omega''(k_{m})&=&1, \quad  g(k_{m})=2 c_0. 
\nonumber 
\end{eqnarray} 
Similarly, at $(\omega, k)=(\omega_m, k_m)$, we find: 
\begin{eqnarray}
\mathbf{R}_b &=& [1,~0,~-1]^T, 
\quad
\mathbf{R}_b'=[0,
~-\frac{\gamma}{\Omega},~0]^T, 
\nonumber \\
\omega''(k_{m})&=&1, \quad  g(k_{m})=2 c_0.
\nonumber
\end{eqnarray}
In this case too, it is clear that $\omega''(k_m) >0$ and $g(k_m)>0$, and hence the system supports a DS solution, given by Eq.~\eqref{eq:dslc} and illustrated 
in Fig.~\ref{fig:dsmb}.  {Notice that we have
 confirmed the absence of unstable eigenvalues for this solution within the realm of the full stability analysis of the BdG equations; once again, 
 see  Appendix~\ref{sec:eig}} for details on the relevant
 BdG computation setup.
 
\subsection{Solitary waves at the upper branch}

\begin{figure}[!htbp]  
\includegraphics[width=\columnwidth]{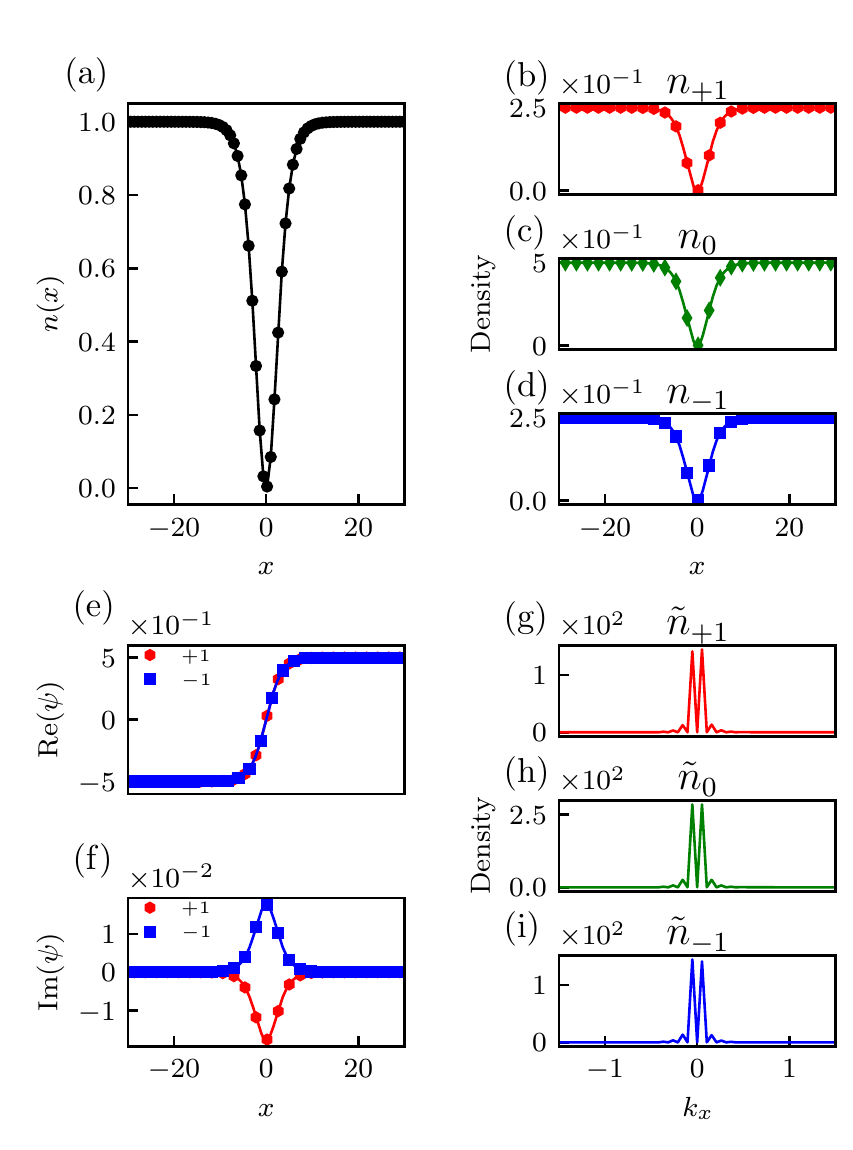}
\caption{The steady state DS solution for the upper branch. The panels depict: (a) The total density, (b)-(d) the density of individual components, (e) the real part of $(\psi_{+1},\psi_{-1})$, (f) the imaginary part of $(\psi_{+1},\psi_{-1})$, and (g)-(i) the density of individual components in Fourier space. Solid line and symbols represent, respectively, numerical results and analytical prediction (Eq.~(\ref{eq:dslc})), with circles representing total density $n$, and squares, diamonds and hexagram denoting spinor components with $m=-1$, $m=0$ and $m=+1$, respectively. The parameters are $c_0=1$, $\Omega=6$, $\gamma=1$, $\lambda_t=0$. $\mu=\omega_m+0.1$, and $C=1$. }
\label{fig:dshb}
\end{figure}

We now consider the upper branch, $\omega_3(k)$, which  
also features a minimum at $k_m=0$.
 In this case, $\omega_m(k_m)= \sqrt{2} \Omega$ (see Fig.~\ref{fig:spectra}) and, correspondingly, we obtain:
\begin{eqnarray}
\mathbf{R}_a &=&[1,~\sqrt{2},~1]^T, \quad \mathbf{R}_a'=\left[\frac{\sqrt{2}\gamma}{\Omega},
~\frac{\gamma}{\Omega},~0\right]^T,
\nonumber \\
 \omega''(k_{m})&=&1+\frac{\gamma^2}{\sqrt{2}\Omega}, \quad 
  g(k_{m})=2 c_0(2+\beta).
\nonumber 
\end{eqnarray}
Similarly, at $(\omega, k)=(\omega_m, k_m)$, we find:
\begin{eqnarray}
\mathbf{R}_b&=&[1,~\sqrt{2},~1]^T, \quad
\mathbf{R}_b'=\left[0,
~-\frac{\gamma}{\Omega},~-\frac{\sqrt{2}\gamma}{\Omega}\right]^T
\nonumber \\ 
\omega''(k_{m}) &=& 1+\frac{\gamma^2}{\sqrt{2}\Omega}, 
\quad  g(k_{m})=2 c_0(2+\beta).
\end{eqnarray}
Obviously, in this setting too, the same sign of $\omega''(k_{m})$ and $g(k_{m})$, indicates the  existence of a DS, which is illustrated in Fig.~\ref{fig:dshb}. In this case as well, the spectral stability of the dark solitary wave has been confirmed by virtue of the BdG analysis.
 
\section{Numerical Results for a trapped condensate}

 In this section we discuss the solitary wave dynamics for a trapped condensate, a system that is naturally of relevance to 
 experiments~\cite{stringari,pethick,frantzeskakis2015defocusing}. We will focus on structures that can be supported in the lower branch of the dispersion relation and discuss representative cases corresponding to the case of a single minimun or two minima (when the lower branch features a double well shape). It is reminded that these cases are distinguished by the relative strength of 
 the SOC parameters $\Omega$ and $\gamma$ ($\gamma^2/\sqrt{2}$, more precisely).
 
\subsection{
Dark solitary waves for \texorpdfstring{$k_m=0$, $\Omega > \gamma^2/\sqrt{2}$}{km=0, Omega > gamma\^2 / sqrt(2)} in the trap
}

 \begin{figure}[!htbp]  
\includegraphics[width=\columnwidth]{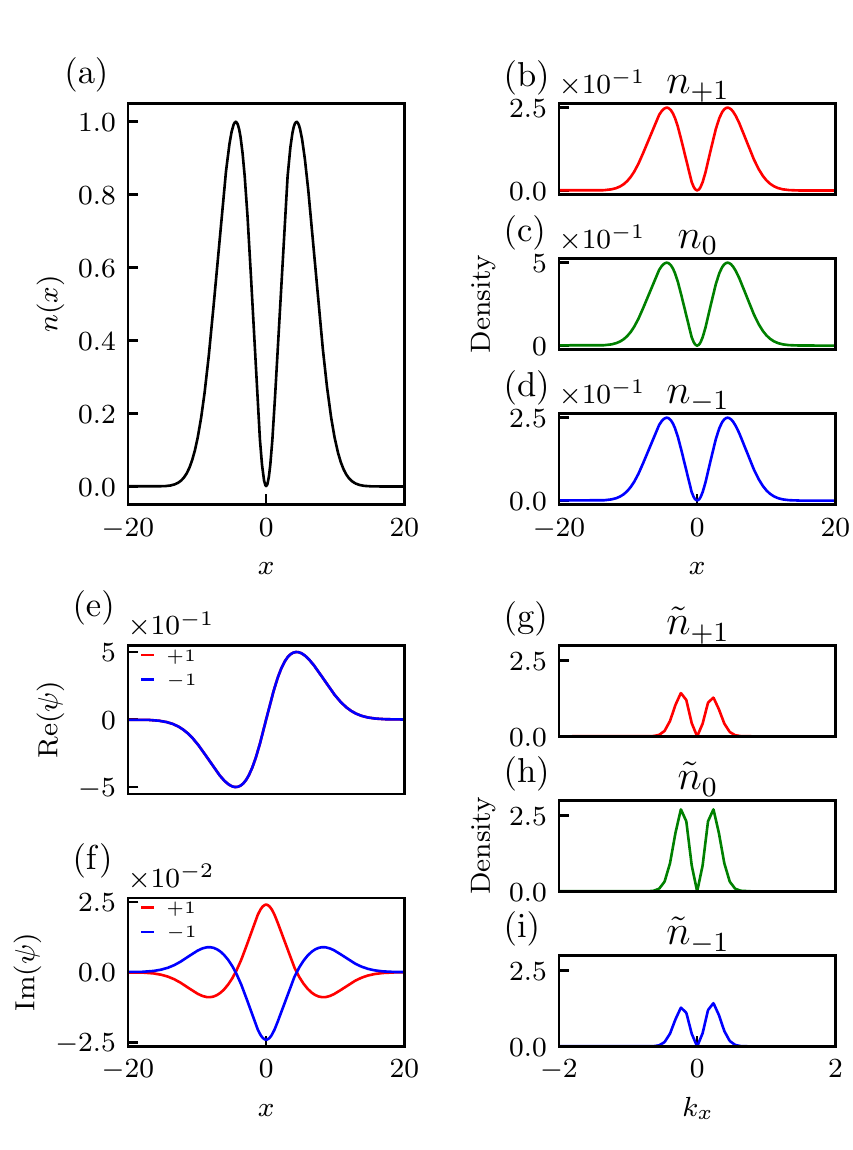}
\caption{The steady state DS solution in the lower branch for $k_m=0$ and  $\Omega>\gamma^2/\sqrt{2}$ in the presence of a trap. 
The panels depict: (a) The total density, (b)-(d) the density of individual components, (e) the real part of $(\psi_{+1},\psi_{-1})$, (f) the imaginary part of $(\psi_{+1},\psi_{-1})$, and (g)-(i) the density of individual components in Fourier space. The parameters are $c_0=1$, $\Omega=6$, $\gamma=1$, $\beta=-0.0046$,  $\lambda_t=0.05$, and $\mu=\omega_m+0.05$.
}
\label{fig:V_O_6_0}
\end{figure}

  \begin{figure}[!htbp]  
\includegraphics[width=\columnwidth]{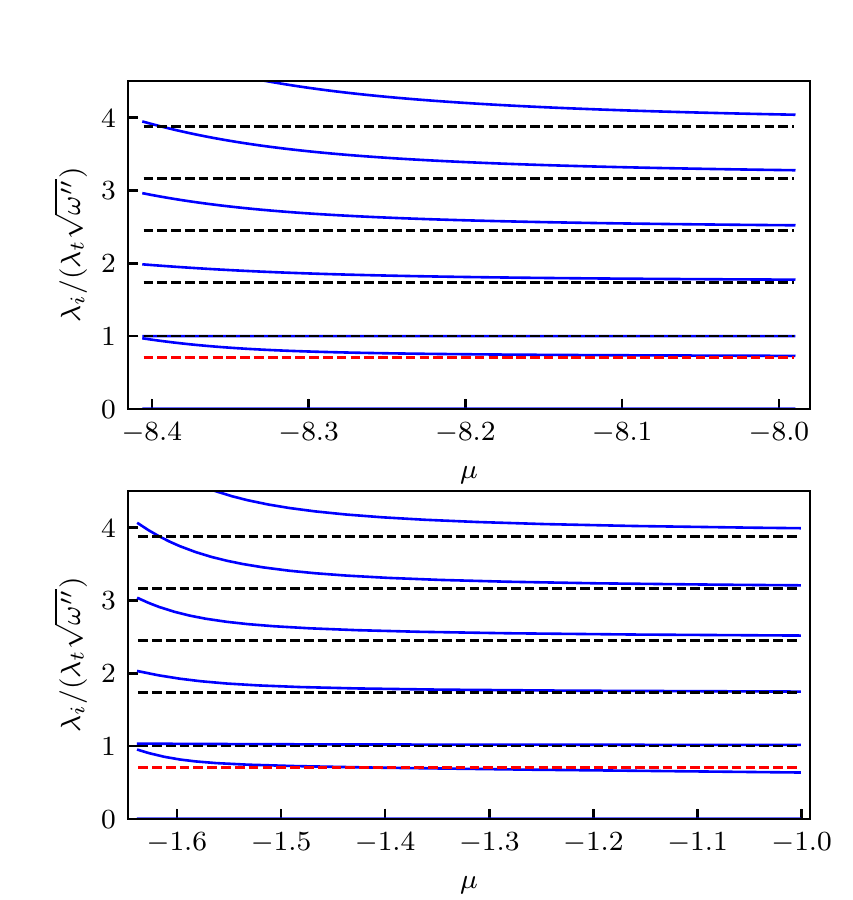}
\caption{The lowest (normalized) imaginary eigenvalues of the dark solitary wave spectrum, as found from the BdG analysis, 
are depicted as functions of $\mu$ 
 for $k_m=0$ and $\Omega>\gamma^2/\sqrt{2}$ by solid blue lines. The parameters are $\Omega=6$ (top panel) and $\Omega=1.2$ (bottom panel), $\gamma=1$, $\beta=-0.0046$ and $\lambda_t=0.05$. 
A comparison with the asymptotic frequency prediction for the DS oscillation frequency
is incorporated as a dashed (red) line, while the remaining asymptotic modes connected to background excitations are depicted by the dashed (black) lines. }
\label{fig:dseigv}
\end{figure}

We start with the steady state solution in the presence of the  trap, shown in Fig.~\ref{fig:V_O_6_0} for $\Omega=6.0$. 
 As before, we identify the stationary state and observe good agreement for each of the components
 with the observed stationary configurations in the figure. Additionally, in this case, the 
 point spectrum of the BdG excitations of the wave in the presence of the parabolic trap~\cite{stringari,frantzeskakis2015defocusing} provides us with an analytical prediction for
 the spectrum of the spinor SOC problem. In order to compare with the numerical BdG spectrum,
 we solve the eigenvalue problem described in Appendix~\ref{sec:eig} to check the spectral stability of the solutions. The way of construction of our stability problem indicates that if the (generally complex) eigenvalue $\lambda$ features a real part, $\lambda_r$, then the pertinent configuration is unstable; on the other hand, if the eigenvalue is imaginary, then the configuration is stable and involves 
 purely oscillatory excitations. 
 
 The lowest imaginary eigenvalues as functions of $\mu$ are shown in Fig.~\ref{fig:dseigv} for $\Omega=6$ (top panel) and $\Omega=1.2$ (bottom panel). The purely imaginary eigenvalues suggest that for large $\mu$, 
 an asymptotic spectral picture is being approached. Within that lies the well-known and extensively studied~\cite{frantzeskakis2015defocusing} mode pertaining to the oscillation of 
 the DS inside the trap 
 of $1/\sqrt{2}$ of the effective trap frequency; the latter, as can be inferred upon rescaling, is found to be equal to $\lambda_t \sqrt{|\omega''(k_m)|}$ in our case. The remaining modes, pertaining to the background (rather than the solitary wave) excitations approach  the values $\sqrt{n(n+1)/2}$ $\lambda_t \sqrt{|\omega''(k_m)|}$~\cite{kevristability2017}. On the other hand, the linear limit (of small enough $\mu$, such that the density tends to vanish) shows that the eigenvalues are integer multiples of the above mentioned effective trap frequency. It is worthwhile to note that as $\Omega$ decreases, we observe a slight
 deviation of the eigenvalues  from the above analytical predictions, although still
 the relevant agreement is fairly reasonable; cf. the 
 bottom panel of Fig.~\ref{fig:dseigv}.

\subsection{Dark solitary waves for 
\texorpdfstring{$k_m=\pm\sqrt{\gamma^4 - 2 \Omega^2}/\gamma$, $\Omega < \gamma^2/\sqrt{2}$}{km=+- sqrt(gamma\^4 - 2 Omega\^2), Omega < gamma\^2 / sqrt(2)}
in the trap}

\begin{figure}[!htbp]  
\includegraphics[width=\columnwidth]{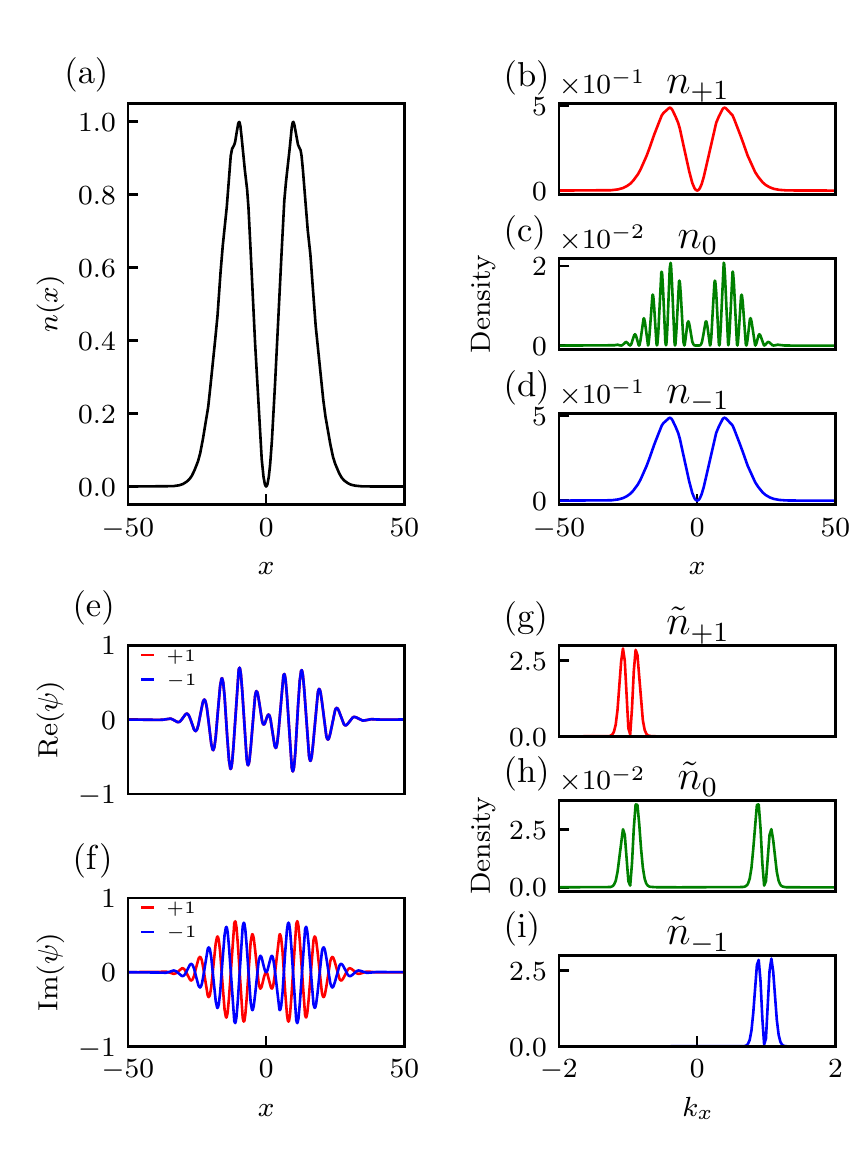}
\caption{The steady state stripe DS solution for 
$k_m=\pm\sqrt{\gamma^4 - 2 \Omega^2}/\gamma$ and 
$\Omega< \gamma^2/\sqrt{2}$ in the presence of the  trap. The panels depict: (a) The total density, (b)-(d) the density of individual components, (e) the real part of $(\psi_{+1},\psi_{-1})$, (f) the imaginary part of $(\psi_{+1},\psi_{-1})$, and (g)-(i) the density of individual components in Fourier space. The parameters are $c_0=1$, $\Omega=0.1$, $\gamma=1$, $\beta=-0.0046$, $\lambda_t=0.01$ and $\mu=\omega_m+0.025$.}
\label{fig:V_O_01_SSDS}
\end{figure}
  \begin{figure}[!htbp]  
 \includegraphics[width=\columnwidth]{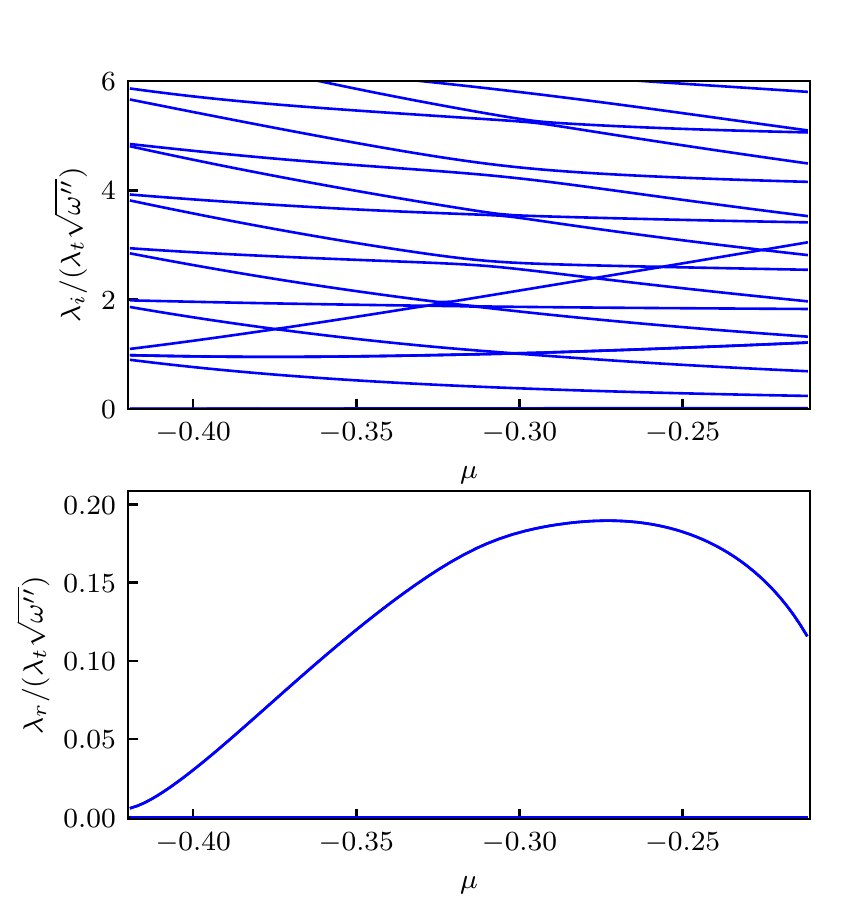}
\caption{The lowest imaginary eigenvalues (top panel) and the lowest real eigenvalues  (bottom panel) of the dark solitary wave spectrum as function of $\mu$ 
for $k_m=\pm\sqrt{\gamma^4 - 2 \Omega^2}/\gamma$ and 
$\Omega<\gamma^2 / \sqrt{2}$ for $\lambda_t=0.05$. 
The parameters are $\Omega=0.1$, $\gamma=1$, $\beta=-0.0046$.
}
\label{fig:dseigvSSDS}
\end{figure}
  \begin{figure}[!htbp]  
  \includegraphics{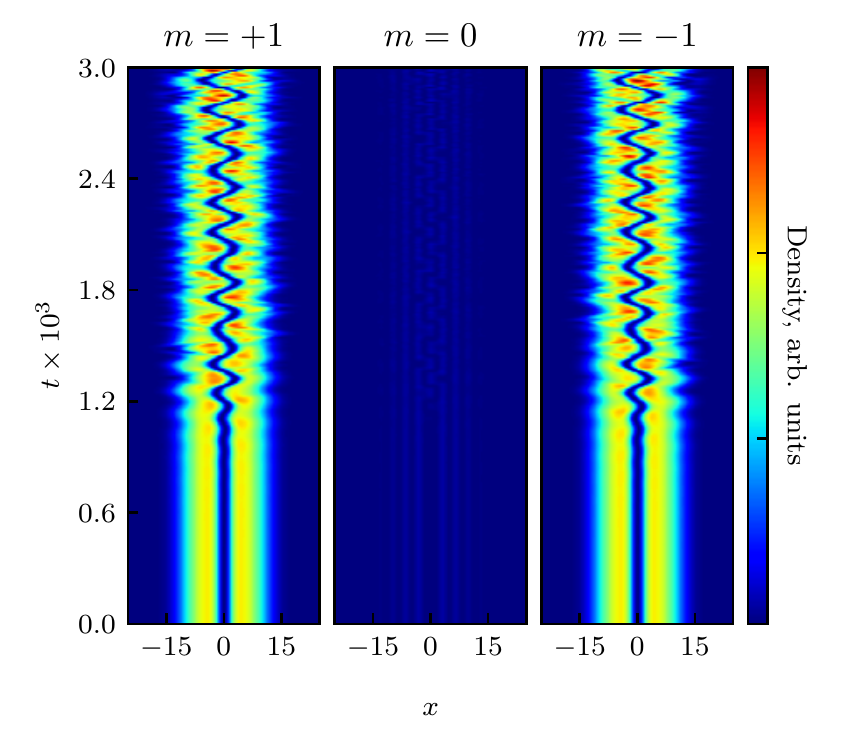}
\caption{Contour plots of the density depicting the evolution of a stripe dark solitary wave perturbed with the eigenvector of the largest real eigenvalue. Shown are the three spin components, $m=+1$ (left), $m=0$ (middle) and $m=-1$ (right). The parameters are $c_0=1$, $\Omega=0.15$, $\gamma=1$, $\beta=-0.0046$, $\lambda_t=0.05$, and $\mu=-0.2730$. A similar behavior  was observed for the case of $\lambda_t=0.01$ (and $\mu=-0.4125$), not shown here.
}
\label{fig:SSDSev}
\end{figure}

In this case, 
our representative example is 
the stripe DS. 
Figure~\ref{fig:V_O_01_SSDS} shows the steady state stripe DS solution in the presence of a trap for $\Omega=0.1$.
It is clear that, despite the confinement  of the relevant state (and its undulations) 
in the parabolic trap, our theory can still adequately capture the relevant configuration. In fact, 
for large $\mu$, this configuration can be approximated by the product of the ground state of the system (in the Thomas-Fermi approximation) and the dark stripe solitary wave that was found in the absence of the trap. 
In Fig.~\ref{fig:dseigvSSDS}, we have also examined the lowest imaginary eigenvalues (top panel) and the lowest real eigenvalues  (bottom panel) as functions of $\mu$ for $\Omega=0.01$. In this case, we have found that, generically, there exist intervals of 
oscillatory instability, as is illustrated in the figure.

Additionally, to study the dynamics of this instability, 
we have perturbed a stripe dark solitary wave with the eigenvector corresponding to the largest real eigenvalue; 
 the evolution of the perturbed stripe DS is shown in Fig.~\ref{fig:SSDSev}. 
 Here, we observe that the oscillatory nature of the instability induces very long-lived oscillations
 of the solitary wave around the center of the trap. Similar results were found also for the case
 of weaker traps, such as  $\lambda_t=0.01$ (not shown here, for brevity).

\section{Results for \texorpdfstring{$\Delta \ne 0$}{Delta != 0}}

In this section, we extend our analysis to $\Delta \ne 0$  (more
specifically setting $\delta_{\rm q}$=0). 
In this case, the dispersion relation, as obtained from the equation $\text{det}(\mathbf{W}) = 0$ [with $\mathbf{W}$ given by Eq.~(\ref{matrixW0})] becomes
\begin{eqnarray}
D(\omega, k) &=& \frac{1}{8}(k^2-2\omega) 
\{[k^2+2(\Delta-\omega)]^2-4\gamma^2 k^2\}
\nonumber \\ 
&-& \Omega^2 [k^2+2(\Delta-\omega)] =0.
\label{matrixW0m}
\end{eqnarray}
In this case, the lowest branch can form a triple-well, in contrast with the $F=1/2$-like double well setting discussed above; i.e., a case bearing three distinct minima for small $\Omega$, as shown in Fig.~\ref{fig:spectra}.
This makes the system very different from a binary SOC-BEC, 
yet it is still experimentally realizable \cite{Valdes-Curiel2021}.
We note, as was shown, e.g., in Ref.~\cite{achilleos2015positive}, that the energy spectrum can be made asymmetric by introducing an energy shift due to a detuning from Raman resonance.


  \begin{figure}[!htbp]  
  \includegraphics[width=\columnwidth]{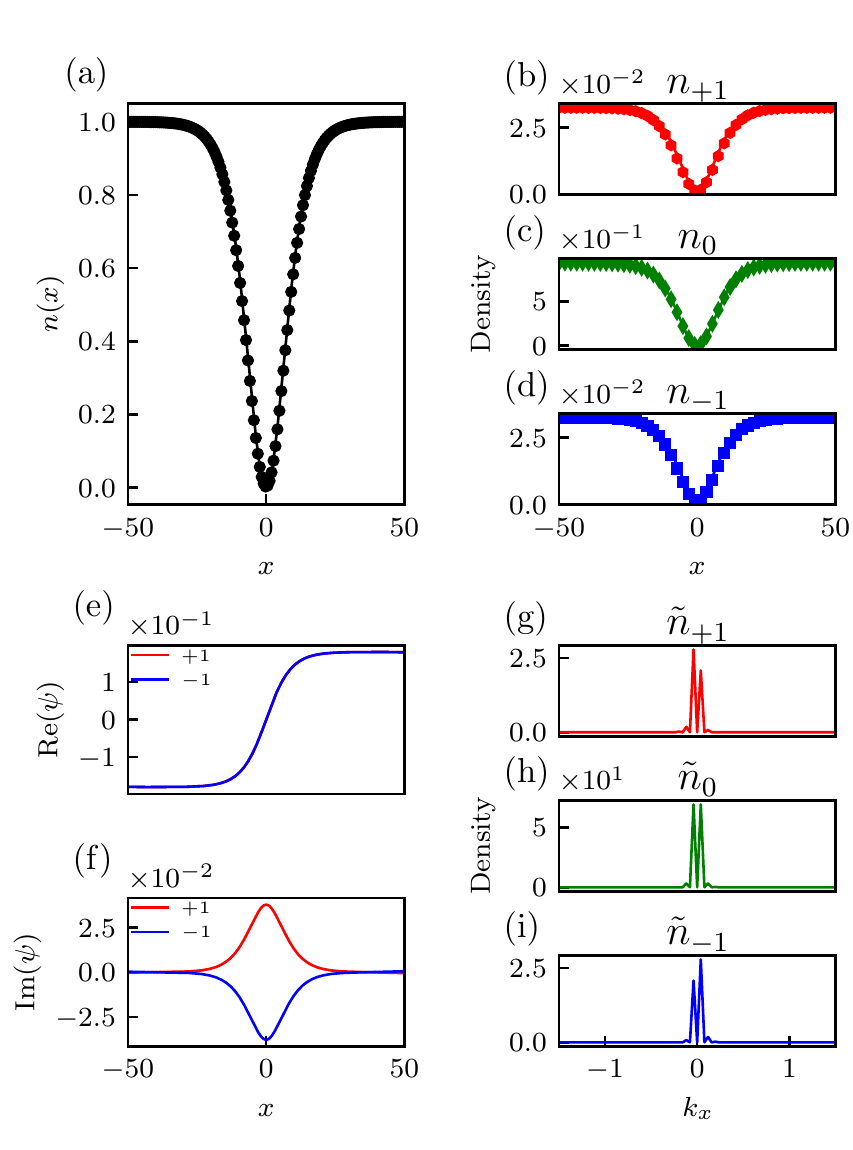}
\caption{The steady state DS solution for $v=0$, $\Omega=0.1$, and $\mu=\omega_m+0.008$. (a) The total density is plotted in (a), and (b)-(d) show the density of individual components. (e)/(f) depicts the real/imaginary part of $(\psi_1,\psi_{-1})$  and (g)-(i) show momentum-density distributions of the individual spin components.
The solid lines represent numerical results and symbols represent the theoretical prediction.}
\label{fig:dsv0momega0_1}
\end{figure}

For non-zero $\Delta$, the eigenfunctions $\mathbf{R_a}=[Q_{1a},Q_{2a},Q_{3a}]^T$ and $\mathbf{R_b}=[Q_{1b},Q_{2b},Q_{3b}]^T$ become
\begin{equation}
\begin{split}
Q_{1a}(\omega,k)&=\frac{\bigg(\frac{k^2}{2}-k\gamma+\Delta-\omega\bigg)}{\bigg(\frac{k^2}{2}+k\gamma+\Delta-\omega\bigg)},\\
Q_{2a}(\omega,k)&=
-\frac{1}{\Omega}\bigg(\frac{k^2}{2}-k\gamma+\Delta-\omega\bigg),~~~
Q_{3a}(\omega,k)=1
\label{eq:REF1b}
\end{split}
\end{equation}
and
\begin{equation}
\begin{split}
 Q_{1b}(\omega,k)&=1,~~Q_{2b}(\omega,k)=
-\frac{1}{\Omega}\bigg(\frac{k^2}{2}+k\gamma+\Delta-\omega\bigg),\\
Q_{3b}(\omega,k)&=1/Q_{1a}(\omega,k).
\label{eq:REF2b}
\end{split}
\end{equation}

Figure~\ref{fig:dsv0momega0_1} shows the analytically and numerically computed solution using $\Omega=0.1$ and $\Delta=\gamma^2/2$ (i.e., quadratic Zeeman shift $\delta_{\rm q}=0$). This solution is obtained for $k_m=0$ of the lowest triple-well band, the single particle ground state. 
As found in Fig.~\ref{fig:dsv0momega0_1}, we expect the $m=0$ component to have the largest contribution to the eigenfunction. 
Moreover, the analytical solution is in a very good agreement with the numerical result.  Another salient feature of the solution is the asymmetric density distribution of $\pm1$ wave functions around the wave vector $k_m=0$ in the Fourier space. This asymmetry is arising from the small but finite linear potential $\pm \gamma k_x$ experienced by the $\pm 1$ wave functions in Fourier space.

 \section{Summary and Future directions}
 
In the present work, we explored the existence, stability, and wherever relevant, dynamics of solitary wave states in $F=1$ SOBECs. 
Although our computations were provided for specific
parameter sets, the methodology used and the structures
considered are expected to be broadly relevant in this
system.
More specifically, we extended the multiscale expansion technique that was applied in two-component systems to analyze the emergence of coherent structures at the extrema of the linear dispersion relation.
We constructed second order approximate solutions, thereby
identifying a wide range of nonlinear excitations including  ``conventional'' dark  and stripe dark solitary waves  as well as bright ones that emerge near the potential maximum of the dispersion relation. 
All of these excitations were corroborated by
means of numerical computations: by first identifying their waveforms via fixed point iterations, and then illustrating their stability via a BdG analysis.

We confirmed their experimental relevance, by studying these states in the presence of a parabolic trap. 
We were able to directly show that the structures persist in confined settings, and to leverage our reduction technique to predict their BdG spectrum in the presence of the trap in good agreement (where appropriate) with direct numerical computations thereof. Here, we have also been able to identify cases where
the trapping may lead to instabilities (e.g. of stripe dark solitary waves)
and have illustrated the corresponding instability-induced dynamics, giving
rise to long-lived DS oscillations.

{The solitary waves herein are particularly interesting when considered from a ``synthetic dimensions'' perspective~\cite{Celi2014} whereby the internal atomic states are assigned a synthetic spatial coordinate.
The Raman coupling introduces an effective magnetic field normal to the plane of  a 2D strip, which is three-site-wide for our $F=1$ case.
This perspective is most useful in the triple-well case where dynamics accurately correspond to the motion of a charged particle in a magnetic field~\cite{Stuhl2015,Mancini2015}.
The anti-symmetry of the momentum distributions Fig.~\ref{fig:dsv0momega0_1}(g) versus (i) therefore implies a type of chiral flow  for this static structure.
The existence of stable traveling solitary waves in this case would lead to dissipationless chiral currents---as in integer quantum Hall systems, but from a completely different mechanism---making the stability of such traveling solitary structures an especially interesting topic for future study. 
Not only is it of relevance to systematically produce such traveling solutions, but this also would pave the way for examining the potential collision of such states and how elastic or inelastic these are. }

More generally, we expect that the provided methodology will
define a playbook for identifying such states in a broad class of spin-orbit coupled systems, including those with the different types of SOC (Rashba or Dresselhaus and combinations thereof) that have been realized in state-of-the-art experiments.
Moreover, our results suggest various near-term research directions.
For instance, it appears natural to consider stationary structures in higher-dimensional systems such as SOC vortices using the corresponding multiscale expansion method and to
explore the associated stability and dynamics.

\acknowledgements

The work of GNK was supported by the Hellenic Foundation for Research 
and Innovation (HFRI) under the HFRI PhD Fellowship grant (Fellowship No.~5860). 
This material is based upon work supported by the US National Science Foundation under Grants DMS-2204702 and PHY-2110030 (P.G.K.).
IBS and ARF were partially supported by the National Institute of Standards and Technology, the National Science Foundation through the Quantum Leap Challenge Institute for Robust Quantum Simulation (OMA-2120757), and the Air Force Office of Scientific Research Multidisciplinary University Research Initiative `RAPSYDY in Q' (FA9550-22-1-0339). 


\appendix
\section{Effective Hamiltonian} {\label{app:Hamiltonian}


\subsection{Bichromatic optical fields}

We begin with a focus on the single-particle term from which we will obtain SOC in a $^{87}$Rb BEC.
Figure~\ref{fig:Setup} depicts our basic setup in which an applied magnetic field ${\bf B} = B_0 \ez$ Zeeman splits the three $m_F$ sub-levels.
We consider the case of large applied magnetic field where, owing to the quadratic Zeeman effect, the energy differences $\delta_{\pm1}$ between $\ket{F=1, m_F = \pm 1}$ and $ \ket{F=1, m_F = 0}$ are significantly different from each other, as indicated.   

In addition a pair of counter propagating laser beams, with equal optical electric field $E_0$ and wavevector $\kr$, drive two photon Raman transitions with strength $\Omega$.
The beam directed along $+\ex$ (red and orange) has two frequency components denoted by $\omega^-_{\pm1}$, while the beam directed along $-\ex$ has a single frequency component $\omega^+$.
As suggested by the level diagram these frequency components will be selected to independently address the $\ket{F=1, m_F = -1}\leftrightarrow \ket{F=1, m_F = 0}$ and $\ket{F=1, m_F = 0}\leftrightarrow \ket{F=1, m_F = +1}$ transitions, as was done experimentally in Ref.~\cite{Campbell2016}.

This combination of laser beams results in the optical electric field 
\begin{align}
{\bf E}({\bf x}) =&\ E_0\Big\{\left[e^{i (\kr x - \omega^-_{-1}t + \pi/2)} + e^{i (\kr x - \omega^-_{+1}t + \pi/2)}\right] \ey \nonumber \\
&\ + e^{-i (\kr x + \omega^+ t + \pi/2)} \ex\Big\},
\end{align}
where the factors of $\pi/2$ serve to establish a convenient spatial origin.
Raman coupling results from the rank-1 tensor (i.e. vector) light shift~\cite{Goldman2014} described by an effective magnetic field
\begin{align}
{\bf B}_{\rm eff}({\bf x}) = &\ i u_v {\bf E}^*({\bf x}) \times {\bf E}({\bf x}) \nonumber \\
=&\ 2 u_v |E_0|^2 \Big[\cos(2\kr x + \delta \omega_{-1} t)  \\
&{\hskip 60 pt} + \cos(2\kr x + \delta \omega_{+1} t) \Big]\ex, \nonumber
\end{align}
in terms of the frequency differences $\delta \omega_{\pm 1} = \omega^+ - \omega_{\pm 1}^-$ and the vector polarizability $u_v$.

This enters into the light-matter Hamiltonian via
\begin{align}
\hat H_{\rm LM} = \frac{g_F \mu_B}{\hbar} {\bf B}_{\rm eff}({\bf x}) \cdot \hat {\bf F},
\end{align}
with Landé $g$-factor $g_F$ and Bohr magneton $\mu_B$, in this case giving a term proportional to $\hat F_x$

\begin{widetext} 
 
\subsection{Rotating wave approximation}

We now eliminate the time-dependence from $\hat H_{\rm LM}$ by first transforming into a rotating frame and then making a pair of rotating wave approximations (RWAs).
In general we consider unitary frame transformations $\hat U(t)$ that take $\ket{\psi^\prime} = \hat U(t) \ket{\psi}$ and recall that $\ket{\psi^\prime}$ evolves according to a rotating frame Hamiltonian
\begin{align}
\hat H_{\rm rot} &= \hat U(t) \hat H \hat U^\dagger(t) - i\hbar \hat U(t) \partial_t \hat U^\dagger(t). \label{eq:rotating_frame}
\end{align}

We consider the unitary frame transformation
\begin{align}
\hat U(t) &= \exp\left[+i\left(\delta\omega_{-1} \ket{-1}\bra{-1} - \delta\omega_{+1}  \ket{+1}\bra{+1}\right) t\right], \label{eq:to_rot_frame}
\end{align}
for which the time-derivative term in Eq.~\eqref{eq:rotating_frame} 
decreases $\ket{m_F = -1}$ in energy by $\hbar \delta \omega_{-1}$ and increases $\ket{m_F = -1}$ in energy by $\hbar \delta \omega_{+1}$, 
The complexity in the problems comes from the remaining operator transform 
\begin{align*}
\hat U(t) \frac{\hat F_x}{\hbar} \hat U^\dagger(t) &= \frac{1}{\sqrt{2}} \left(e^{-i \delta\omega_{-1} t} \ket{0}\bra{-1} +  e^{-i \delta\omega_{+1} t} \ket{+1}\bra{0} + \rm{H.c}.\right),
\end{align*}
leading to a total of 16 terms in the light matter Hamiltonian
\begin{align}
\hat H_{\rm LM} =&\ \frac{g_F \mu_B u_v |E_0|^2}{\sqrt{2}} \Big\{ \label{eq:RotatingFrame}\\
& \Big[e^{2 i \kr x} + e^{-i(2\kr x + 2 \delta\omega_{-1} t)}  + e^{i(2\kr x + (\delta\omega_{+1} + \delta\omega_{-1}) t)} + e^{-i(2\kr x + (\delta\omega_{+1} + \delta\omega_{-1}) t)}\Big] \ket{0}\bra{-1} \nonumber \\
&\ + \Big[e^{i(2\kr x + (\delta\omega_{-1} - \delta\omega_{+1})} + e^{-i(2\kr x + (\delta\omega_{-1} + \delta\omega_{+1})} + e^{2 i \kr x} - e^{-i(2\kr x + 2 \delta\omega_{+1} t)}\Big] \ket{+1}\bra{0} \nonumber\\
&\ + {\rm H.c.}\Big\} \nonumber
\end{align}


The rotating wave approximation consists of eliminating all rapidly rotating terms leading to the final RWA Hamiltonian
\begin{align}
\hat H_{\rm RWA} &= \delta_q \big(\ket{-1}\bra{-1} + \ket{+1}\bra{+1}\big) +  \Omega_R \left[ e^{2 i \kr x} \ket{0}\bra{-1} +  e^{2 i \kr x} \ket{+1}\bra{0} + {\rm H.c} \right], \label{eq:RWA}
\end{align}
where we aggregated the numerical prefactors in Eq.~\eqref{eq:RotatingFrame} into the Raman coupling strength $\Omega_R$ and introduced
$\delta_q= \delta_{-1} - \delta\omega_{-1} = \delta_{+1} - \delta\omega_{+1}$ (thereby constraining $\omega_{\pm1}$).

In practice, terms rotating more rapidly than $\approx 100\ {\rm kHz}$ can be safely neglected as they exceed both the single particle and interaction energy scales in the problem.
For the $^{87}$Rb system specifically this implies that $B_0 \gtrsim 30\ {\rm G}$ so that both the linear and quadratic Zeeman shifts are above this scale.

This leads to the single particle Hamiltonian
\begin{equation}
\begin{split}
\hat{H_0}&=\frac{\hbar^2 k^2}{2\ma} + \frac{\Omega}{\sqrt{2} }  \big[\cos(2k_R x)\hat F_x- \sin(2k_R x) \hat F_y\big]+\frac{\delta_{\rm q}}{\hbar} F_z^2).
\label{eq:APP_H1}
\end{split}
\end{equation}
A final spin rotation about $\mathbf{e}_z$ by an angle $2 k_R x$ leads to the spin-orbit coupled Hamiltonian
\begin{equation}
\hat{H_0}=\frac{(-i\hbar\partial_x \hat I + k_R \hat F_z)^2}{2\ma}+ \frac{\delta_{\rm q}}{\hbar} \hat F_z^2 +  \frac{\Omega}{\sqrt{2} }  \hat F_x,
\label{eq:APP_H3}
\end{equation}
in Eq.~\eqref{eq:H0}.

\subsection{Interaction Hamiltonian}

We now turn our attention to the four field terms in Eq.~\eqref{eq:second_quantized_H}.
In the second quantized notation, the transformation analogous to Eq.~\eqref{eq:to_rot_frame} is
\begin{align}
\hat {\mathcal U}(t) &= \exp\left[+i\left(\delta\omega_{-1} \hat n_{-1}(x) - \delta\omega_{+1}  \hat n_{+1}(x)\right) t\right], 
\end{align}
for example giving the parallel action of $\hat U(t) \ket{+1} = e^{-i \delta\omega_{+1} t} \ket{+1}$ versus $\hat {\mathcal U}(t) \hat \psi_{+1}(x) \hat {\mathcal U}^\dagger(t) = e^{-i \delta\omega_{+1} t} \hat \psi_{+1}(x)$.
We now consider the action of this transformation on the interaction contribution to the many-body Hamiltonian

\begin{align}
\hat {\mathcal H}_{\rm int} =\frac{1}{2} \int d x:\left[ g_0 \hat n^2(x)+ \frac{g_2}{ \hbar^2}\left| \hat{\boldsymbol{\mathcal F}}(x) \right|^2 \nonumber
\right]:
\end{align}
where $:\cdots:$ denotes the normal ordering operation.
We now consider the term-by-term action of our rotation on this many-body Hamiltonian.
The total density is trivially unchanged
\begin{align}
\hat {\mathcal U}(t) :\hat{n}^2(x): \hat {\mathcal U}^\dagger(t) = :\hat{n}^2(x):
\end{align}
and acquires no time dependence.
By contrast the spin dependent term
\begin{align}
\hat {\mathcal U}(t) :\left| \hat{\boldsymbol{\mathcal F}}(x) \right|^2:  \hat {\mathcal U}^\dagger(t)& = \hat {\mathcal U}(t) \Big[ \hat \psi_{+1}^\dagger \hat \psi_{+1}^\dagger \hat \psi_{+1} \hat \psi_{+1} + \hat \psi_{-1}^\dagger \hat \psi_{-1}^\dagger \hat \psi_{-1} \hat \psi_{-1}+2 \hat \psi_{+1}^\dagger \hat \psi_{0}^\dagger \hat \psi_{+1} \hat \psi_{0}+2 \hat \psi_{-1}^\dagger \hat \psi_{0}^\dagger \hat \psi_{-1} \hat \psi_{0}-2\hat \psi_{+1}^\dagger \hat \psi_{-1}^\dagger \hat \psi_{+1} \hat \psi_{-1}  \nonumber \\
&\ +2 \hat \psi_{0}^\dagger \hat \psi_{0}^\dagger \hat \psi_{+1} \hat \psi_{-1}+2 \hat \psi_{+1}^\dagger \hat \psi_{-1}^\dagger \hat \psi_{0} \hat \psi_{0} \Big]  {\mathcal U}^\dagger(t) \\ \nonumber 
& = \Big[ \hat \psi_{+1}^\dagger \hat \psi_{+1}^\dagger \hat \psi_{+1} \hat \psi_{+1} + \hat \psi_{-1}^\dagger \hat \psi_{-1}^\dagger \hat \psi_{-1} \hat \psi_{-1}+2 \hat \psi_{+1}^\dagger \hat \psi_{0}^\dagger \hat \psi_{+1} \hat \psi_{0}+2 \hat \psi_{-1}^\dagger \hat \psi_{0}^\dagger \hat \psi_{-1} \hat \psi_{0}-2\hat \psi_{+1}^\dagger \hat \psi_{-1}^\dagger \hat \psi_{+1} \hat \psi_{-1}  \nonumber \\
&\ +2 \hat \psi_{0}^\dagger \hat \psi_{0}^\dagger \hat \psi_{+1} \hat \psi_{-1} e^{-i (\delta\omega_{+1}-\delta\omega_{-1}) t} +2 \hat \psi_{+1}^\dagger \hat \psi_{-1}^\dagger \hat \psi_{0} \hat \psi_{0} e^{i (\delta\omega_{+1}-\delta\omega_{-1}) t} \Big]  \nonumber
\end{align}
does have time-dependent contributions in the spin-changing collision terms, which are the eliminated by the RWA.
This leads to the RWA expression used in main manuscript
\begin{align}
:\left| \hat{\boldsymbol{\mathcal F}}_{\rm RWA}(x) \right|^2: & \equiv \left[ \hat \psi_{+1}^\dagger \hat \psi_{+1}^\dagger \hat \psi_{+1} \hat \psi_{+1} + \hat \psi_{-1}^\dagger \hat \psi_{-1}^\dagger \hat \psi_{-1} \hat \psi_{-1}+2 \hat \psi_{-1}^\dagger \hat \psi_{0}^\dagger \hat \psi_{-1} \hat \psi_{0}+2 \hat \psi_{+1}^\dagger \hat \psi_{0}^\dagger \hat \psi_{+1} \hat \psi_{0}-2\hat \psi_{+1}^\dagger \hat \psi_{-1}^\dagger \hat \psi_{+1} \hat \psi_{-1}\right].
\end{align}

\end{widetext} 

\section{Effective mass} \label{meff}
As was shown in Sections~III.B and IV.A, the effective NLS equation 
\begin{equation}
i\varphi_T+\frac{1}{2} \omega''(k_m) \partial_X^2 \varphi  -g(k_m)|\varphi|^2 \varphi + \omega_0 \varphi=0,
\label{NLSb}
\end{equation}
can have negative prefactors for both dispersion and nonlinearity, i.e., $\omega''(k_m)=-|\omega''(k_m)|<0$ and $-g(k_m)<0$. 
In this case, the NLS~(\ref{NLSb}) possesses a stationary bright solitary wave solution~(\ref{eq:bs}).
Starting from this stationary waveform, one may use the Galilean invariance of the NLS equation, and construct a traveling bright solitary wave of the form 
\begin{eqnarray}
\varphi_{BS}(X,T)&=&\sqrt{\frac{2 \omega_0}{g(k_m)}} 
\text{sech}\Bigg[\sqrt{\frac{2 \omega_0}{|\omega''(k_m)|}}(X-v_s T)\Bigg]
\nonumber \\
&&\times \exp{i(k_s X-\omega_s T)}.
\label{tsw}
\end{eqnarray}
The above solitary wave is characterized by a velocity $v_s$, a frequency $\omega_s$ and a wavenumber $k_s$ that are connected by a ``solitary wave dispersion relation''
\begin{equation}
v_s=-|\omega''(k_m)| k_s, \quad 
\omega_s=-\frac{1}{2} |\omega''(k_m)| k_s^2. 
\end{equation}
Notice that the velocity $v_s$ can be directly obtained from the dispersion relation using $v_s = \partial \omega_s/\partial k_s = -|\omega''(k_m)| k_s$. 
It is straightforward to find that the NLS Eq.~(\ref{NLSb}) conserves the momentum $P$ (i.e., $\partial P/\partial T =0$), which is given by:
\begin{equation}
P=\frac{i}{2}\frac{g(k_m)}{|\omega''(k_m)|}\int_{-\infty}^{\infty}\left(\varphi \varphi_X^{\ast}-\varphi^{\ast} \varphi_X \right) dX.
\label{mom}
\end{equation}
Substituting the traveling solitary wave~(\ref{tsw}) into Eq.~(\ref{mom}), one finds the solitary wave momentum
\begin{equation}
P_s = 2\sqrt{2\omega_0} k_s.
\label{Ps}
\end{equation}
Leveraging the particle picture of a solitary wave~(\ref{tsw}), we determine the effective mass $m_{\rm eff}$ from
\begin{equation}
m_{\rm eff} = \frac{\partial P_s}{\partial v_s},  
\label{msol1}
\end{equation}
which leads to
\begin{equation}
m_{\rm eff} =\frac{\partial P_s/\partial k_s}{\partial v_s/\partial k_s} = -\frac{2\sqrt{2\omega_0}}{|\omega''(k_m)|}.
\label{msol2}
\end{equation}
As mentioned in Section~IV, the bright solitary wave, which exists for $\omega''(k_m)<0$, features a negative effective mass.
Contrary, as suggested by Eq.~(\ref{msol2}), dark solitary waves which exist for $\omega''(k_m)>0$, have positive effective mass.

\section{Linearized GPE and BdG analysis} \label{sec:eig}

The dimensionless coupled GPE equations can be expressed as
\begin{equation}
  i\partial_t \mathbf{\Psi} = \left(\hat{h}+c_0 n \mathbf{I} +c_2 \mathbf{A}_1 \right) 
  \mathbf{\Psi},
  \label{a1}
\end{equation}
with
\begin{align*}
\hat{h} &=\frac{1}{2}\left(-\partial^2_{x} \mathbf{I}-i \gamma \partial_{x}  f_z\right)+V(x)\mathbf{I}+\sqrt{2}\Omega f_x,
\end{align*}
and 
\begin{align*}
\mathbf{A}_1&=
  \begin{bmatrix}
    n_1+n_0-n_{-1} &        0   &   0  \\
         0         & n_1+n_{-1} &   0  \\
         0         &     0      &  n_{-1}+n_0-n_1 
  \end{bmatrix},
\end{align*}
in terms of the densities $n_i=|\psi_i|^2=\psi_i^{\ast}\psi_i$.
Then, letting $\mathbf{\Psi}_0=(\psi_{+1},\psi_0,\psi_{-1})$ be a steady state solution, we consider small perturbations $\delta \mathbf{\Psi}$ around the steady state, and introduce the ansatz 
\begin{align*}
\mathbf{\Psi} &=(\mathbf{\Psi}_0 +\epsilon\,
\delta \mathbf{\Psi})\exp(-i \mu t),
\end{align*}
into Eq.~(\ref{a1}).
At order $O(\epsilon)$, we obtain the linear equation
\begin{equation}
  i\partial_t \delta \mathbf{\Psi} = \mathbf{K}_R\, \delta \mathbf{\Psi}+ \mathbf{K}_I\, \delta\mathbf{\Psi}^{\ast},
\label{a2}
\end{equation}
for $\delta \mathbf{\Psi}$, where
\begin{eqnarray}
  \mathbf{K}_R&=&\hat{h}+c_0 n \mathbf{I}-\mu+ c_0 \mathbf{B}_1 +c_2 \mathbf{C}_1,
\notag
\\
\notag
  \mathbf{K}_I&=&\hat{h}+c_0 n \mathbf{I}+c_0 \mathbf{B}_2+ c_2 \mathbf{C}_2,
\end{eqnarray}
and
\[
\mathbf{B}_1=
  \begin{bmatrix}
     n_1 &  \psi_0^{\ast}\psi_{+1}  &  \psi_{-1}^{\ast}\psi_{+1}  \\
     \psi_{+1}^{\ast}\psi_0 &  n_0&  \psi_{-1}^{\ast}\psi_0\\
     \psi_{+1}^{\ast}\psi_{-1} &  \psi_{0}^{\ast}\psi_{-1} &  n_{-1}
  \end{bmatrix}
,
\]
\[
\mathbf{C}_1=
  \begin{bmatrix}
    F_z+n_1+n_0 & \psi_{0}^{\ast}\psi_{+1} & -\psi_{-1}^{\ast}\psi_{+1} \\
    \psi_{+1}^{\ast}\psi_{0} & n_{-1}+n_1& \psi_{-1}^{\ast}\psi_{0} \\
   - \psi_{+1}^{\ast}\psi_{-1}  & \psi_{0}^{\ast}\psi_{-1}  & -F_z+n_{-1}+n_0
  \end{bmatrix}
,
\]
\[
\mathbf{B}_2=
  \begin{bmatrix}
    \psi_{+1}^{2} & \psi_{0} \psi_{+1} & \psi_{-1}\psi_{+1} \\
    \psi_{+1} \psi_{0} &  \psi_{0}^2 & \psi_{-1} \psi_{0} \\
    \psi_{+1} \psi_{-1}  & \psi_{0} \psi_{-1}  & \psi_{-1}^{2}
  \end{bmatrix}
,
\]
\[
\mathbf{C}_2=
  \begin{bmatrix}
    \psi_{+1}^{2} & \psi_{0} \psi_{+1} & -\psi_{-1}\psi_{+1} \\
    \psi_{+1} \psi_{0} &  0 & \psi_{-1} \psi_{0} \\
   -\psi_{+1} \psi_{-1}  & \psi_{0} \psi_{-1}  & \psi_{-1}^{2}
  \end{bmatrix}
,
\]
Finally, inserting a perturbation of the form
$$\delta \mathbf{\Psi} = \mathbf{P} \exp(\lambda t)+ \mathbf{Q}^{\ast} \exp(\lambda^{\ast} t),$$ into the
the linearized problem~(\ref{a2}) gives the coupled equations
\begin{equation}
\notag
\left\{
\begin{array}{rcl}
i\lambda \mathbf{P} &=& \mathbf{K}_R \mathbf{P} 
+\mathbf{K}_I \mathbf{Q}
\\[1.0ex]
i\lambda \mathbf{Q} &=& - \mathbf{K}_R^{\ast} \mathbf{Q} - \mathbf{K}_I^{\ast} \mathbf{P}
\end{array}
\right.
\end{equation}
which can be explicitly written as the eigenvalue problem
\begin{equation}
\label{eq:evalpb}
\mathbf{M}\mathbf{V} = \lambda \mathbf{V},
\end{equation}
with
$$\mathbf{M}=
-i
\begin{bmatrix}
\mathbf{K}_R & \mathbf{K}_I\\
-\mathbf{K}_I^{\ast} & -\mathbf{K}_R^{\ast}
\end{bmatrix}
\quad
\text{and}
\quad
\mathbf{V}=
\begin{bmatrix}
\mathbf{P}\\
\mathbf{Q}
\end{bmatrix}. 
$$

\bibliographystyle{apsrev4}
\let\itshape\upshape
\normalem
\bibliography{output}

\end{document}